\documentclass[11pt]{article}

\usepackage[letterpaper,margin=1in]{geometry} % use a4paper if needed

\usepackage[english]{babel}

\usepackage[
  style=apa,
  bibstyle=authoryear,
  autocite=inline,
  maxcitenames=2,
 maxbibnames=12,
  backend=biber
]{biblatex}
\usepackage{setspace}
\usepackage[utf8]{inputenc}
\usepackage[T1]{fontenc}
\usepackage{csquotes}  
\usepackage{graphicx}
\usepackage[export]{adjustbox}
\graphicspath{ {./images/} }

\usepackage{amsmath,amssymb}
\usepackage{mathabx,mathrsfs}
\usepackage[version=4]{mhchem}
\usepackage{stmaryrd}
\usepackage{bbold}
\usepackage{tikz}
\usepackage{indentfirst}
\usepackage{caption}
\usepackage{float}
\usepackage{hyperref}
\usepackage{subcaption}
\usepackage{algorithm}
\usepackage{algpseudocode}
\usepackage{mathtools}
\usepackage{array}
\usepackage{blindtext}
\usepackage{titlesec}
\usepackage{xcolor}
\usepackage{multirow}
\usepackage[final]{microtype} % better spacing & hyphenation
\definecolor{darkblue}{HTML}{0B2F6A}
\usepackage{amsthm} % Required for the theorem environment

\usepackage[normalem]{ulem}

\hypersetup{colorlinks,linkcolor={blue},citecolor={blue},urlcolor={red}}

\newtheorem{remark}{Remark}

\DeclareMathOperator*{\argmax}{arg\ max}

\DeclareMathOperator*{\argmin}{arg\ min}
\DeclareMathOperator*{\sign}{sgn}

\newcommand{\EE}{\mathbb{E}}

\newcommand{\bc}{\mathbf{c}}
\newcommand{\bx}{\mathbf{x}}

\newcommand{\bz}{\mathbf{z}}
\newcommand{\bC}{\mathbf{C}}
\newcommand{\bI}{\mathbf{I}}

\def\cD{\mathcal{D}}

\def\cF{\mathcal{F}}

\def\cJ{\mathcal{J}}
\def\cK{\mathcal{K}}

\def\cM{\mathcal{M}}
\def\cN{\mathcal{N}}

\def\cP{\mathcal{P}}

\def\cS{\mathcal{S}}

\def\cX{\mathcal{X}}

\def\bC{\mathbf{C}} % bold

\def\bE{\mathbb{E}}

\def\bI{\mathbf{I}} % bold

\title{
Battery Storage Co-Optimization in Day-Ahead and Real-Time Markets with Bayesian Optimization
}

\author{
Thiha Aung\thanks{
Department of Statistics and Applied Probability,
University of California, Santa Barbara.
Email: taung@ucsb.edu.
}
\and
Mike Ludkovski\thanks{
Department of Statistics and Applied Probability,
University of California, Santa Barbara.
Email: ludkovski@pstat.ucsb.edu.
}
}

\date{\today}

\begin{document}

\maketitle

\begin{abstract}
We propose Adaptive Refinement Bayesian Optimization for 
Day-Ahead and Real-Time (ARBO-DART) markets, an algorithm for BESS intraday dispatch
co-optimization in which day-ahead  (DA) commitment profiles are optimized against 
value of real-time (RT) recourse computed by a black-box stochastic control solver. In our framework, the DA price curve is taken as exogenous and RT prices evolve as a mean-reverting process around it. The RT recourse layer performs dynamic closed-loop control while accounting for the piecewise-linear state-of-charge dynamics and the DA-driven feasible control set. By wrapping Bayesian Optimization (BO) around the RT solver, ARBO-DART jointly optimizes DA commitments and dynamic RT flexibility without requiring analytic gradients, closed-form value functions, or finite-scenario approximations. To overcome the curse of dimensionality in fixed-resolution DA profiles,
ARBO-DART starts from a coarse partition of DA commitments and
progressively refines charge and discharge blocks where additional temporal
resolution is most needed, as judged by the corresponding RT policy.  
Numerical experiments across realistic DA price curves reveal the effectiveness
of ARBO-DART in recovering economically meaningful DA bidding structures while being several times faster relative to fixed-resolution
comparators.
\end{abstract}

\section{Introduction}

Grid-scale battery energy storage systems (BESS) are being deployed at record levels,
reflecting both declining storage costs and growing demand for flexible resources \parencite{WoodMacACP2026StorageMonitor,ColeKarmakar2025BatteryCostProjections}. 
As more and more batteries enter wholesale markets, profitability is increasingly determined not only by deployment scale but by how effectively operators monetize market fluctuations day-to-day. 

Economic viability of BESS depends on monetizing dispatch flexibility across several revenue streams, including energy arbitrage \parencite{jaber2024tradingpropagatorsconstraintsapplications,sandbergen_optimized_2025,sage_enhancing_2024,DART_SP_Krish,Focker_Planck_wind_BESS}, ancillary services (AS) such as frequency regulation and reserves \parencite{Shi_solo_bessfreq_reg,mishra_hybrid_bessfreq_reg}, capacity or resource-adequacy payments, and stacked-service provision across multiple products \parencite{Tong2018EnergySS,BESS_stacked_services_review}. As AS markets saturate, energy arbitrage is becoming increasingly central to BESS profitability. For example,  recent market evidence in the Texas ERCOT grid shows that rising BESS offers into AS have contributed to lower clearing prices, shifting the battery revenue stack toward energy arbitrage \parencite{ModoERCOTAS2025,ModoBESSRevenue2026}. Similar patterns have been observed in CAISO, where energy arbitrage constitutes a dominant share of merchant BESS revenues \parencite{ModoBESSRevenue2026,CAISO2025BatteryStorageReport}. These trends make capturing intraday price fluctuations across day-ahead (DA) and real-time (RT) electricity markets ever more critical, motivating a growing literature on battery valuation frameworks \parencite{Bolun_Xu_valuation,lohndorf_value_2023,hornek_value_2025,cogneville_battery_2024}.

Driven by such trends, our paper focuses on joint energy arbitrage by standalone BESS across DA and RT (DART) wholesale electricity markets. The DA market allows storage operators to lock in intertemporal price spreads before the operating day, while the RT market exposes them to high-frequency uncertainty and creates opportunities for dynamic adjustments.  The joint setting is challenging because battery dispatch is inherently intertemporal: charge and discharge decisions are coupled throughout the day through the state-of-charge (SoC) and are constrained by power and energy limits. This coupling extends across market layers as DA commitments determine the remaining SoC flexibility available for RT recourse. Indeed, 
\textcite{goutte2019value} show that substantial BESS value is unlocked by augmenting hourly DA decisions with sub-hourly RT adjustments. Thus, BESS operation requires a \emph{co-optimization} framework that jointly accounts for DA bidding decisions and closed-loop RT control.

Much of the existing literature studies energy storage dispatch in a single market layer. One stream focuses on DA bidding and scheduling \parencite{fleten_short-term_2008,faria_day-ahead_2011,lohndorf_optimizing_2013,hinz_optimal_2017}, while another studies intraday financial trading in the spot or forward markets \parencite{jiang_optimal_2015,aid_optimal_2015, 
bertrand_adaptive_2020, 
hornek_value_2025,cogneville_battery_2024,schaurecker_maximizing_2025}. 

The simplest approach to consider both DA and RT decisions is sequential optimization, in which the DA profile is chosen first and the RT policy is optimized afterward, conditional on the DA decisions \parencite{sandbergen_optimized_2025,DART_SP_Krish}. While computationally attractive, this approach is suboptimal because the DA profile is not chosen by directly optimizing the downstream value of RT flexibility.
\textcite{boomsma_bidding_2014} compare sequential and coordinated bidding in DA and balancing markets using multi-stage stochastic programming and show that coordinated optimization outperforms sequential bidding.  Existing DART co-optimization approaches differ in how they represent uncertainty and recourse. Deterministic formulations optimize DA and RT decisions under fixed power price trajectories and provide tractable benchmark models \parencite{rolling_intrinsic_MILP}. Scenario-based stochastic programming improves on this by representing RT uncertainty through a finite set of price scenarios, enabling coordinated DA and RT decisions \parencite{boomsma_bidding_2014,finnah_integrated_2022,lohndorf_value_2023,Kim_two_stage_StochOpt}. When the problem has suitable convex or linear structure, decomposition methods, such as stochastic dual dynamic programming can further exploit the temporal structure of multi-stage decisions. However, these approaches become difficult to apply when the RT layer is nonlinear, path-dependent, and high-dimensional.
Recent approaches have therefore considered more flexible RT solvers, including reinforcement learning and regression Monte Carlo methods \parencite{RL_stochastic_DA_RT_transformer,liu2025cooptimization,Aung_Ludkovski_CDC}.

A related stream of work studies statistical modeling and forecasting of DART spreads themselves. \textcite{GalarneauVincentEtAl2023DARTSpikes} develop statistical-learning models to predict DART spread spikes and show that spike-probability forecasts can be used as profitable trading signals. \textcite{ForgettaEtAl2025DARTMixture} propose a covariate-dependent mixture model for the distribution of hourly DART spreads, distinguishing regular spread behavior from positive and negative spike regimes.  \textcite{HubertLolasSircar2026TradingElectrons} study DART spread spike prediction across ISO electricity markets and combine directional spread forecasts with trading decisions. These papers highlight the statistical complexity of DART spreads as a market object. Our focus is complementary: we embed DART price uncertainty into a physical BESS co-optimization problem with SoC dynamics, operational constraints, and RT recourse, keeping our framework agnostic regarding DART dynamics.

\paragraph{Summary of contributions.} 
We develop a DART \emph{co-optimization} framework for standalone BESS in which the RT recourse layer is accessed through a closed-loop stochastic control solver. Our objective is motivated by the two-stage stochastic optimization formulation of \textcite{Kim_two_stage_StochOpt}, which optimizes DA and RT decisions using a finite-scenario stochastic program solved by conventional optimization methods. In contrast, in our framework the RT recourse value is supplied by a richer black-box solver, which can handle uncertainty representations and recourse dynamics beyond finite scenarios.

Our contributions are threefold. First, we formulate the RT recourse problem as a stochastic control problem whose feasible action and state spaces are conditional on the fixed DA profile. In our implementation, RT prices are modeled as mean-reverting around the DA price trajectory, while the BESS state evolves according to piecewise-linear SoC dynamics with power and SoC constraints. The RT objective combines linear payoff, a quadratic trading friction on RT recourse, and a terminal SoC penalty. Building on the SHADOw-GP battery control solver developed in \textcite{Aung_Ludkovski_CDC}, we adapt this method to the DART setting, where the feasible RT policy space is jointly determined by DA commitments, battery power limits, and SoC constraints. Unlike reinforcement learning approaches that often enforce battery constraints through reward penalties or heuristic tuning \parencite{SAGE_RL_review}, our approach handles power and state constraints directly through feasible-control projections.

Second, we introduce a BO framework for DART co-optimization. BO is well suited for expensive black-box objectives, where sequential surrogate modeling and acquisition functions are used to allocate evaluations efficiently \parencite{jones1998efficient,shahriari2016taking,frazier2018tutorial}. In our setting, the black-box objective is not a static simulator or hyperparameter-tuning loss, but the value of a DA battery profile after solving the induced closed-loop RT stochastic control problem. BO therefore adaptively searches over DA profiles, with each candidate profile evaluated through the corresponding RT recourse solver. This enables co-optimization of DA profiles and RT flexibility without requiring analytic gradients, finite-scenario recourse approximations, or a closed-form representation of the RT value function. To the best of our knowledge, this is the first use of BO for a  co-optimization that combines static and dynamic control, a strategy that would be applicable in many other diverse contexts.

Third, we develop an adaptive refinement procedure for the DA profiles. A brute-force BO formulation is inefficient and may lead to solutions that are not economically meaningful. 
Instead, we represent DA dispatch using a block structure that is iteratively refined. The resulting ARBO-DART algorithm starts from a coarse partition and adaptively adds and prunes DA blocks based on the corresponding RT recourse. This approach self-learns the effective dimension of the outer DA problem, providing a more parsimonious DA profile. By keeping the outer BO problem lower-dimensional, we show that ARBO-DART runs 2-4 times faster that the base full-dimension BO. 

The rest of the paper is organized as follows. Section~\ref{sec:problem_setup} introduces the price models, BESS dynamics, and DART optimization objective. Section~\ref{sec:rt_solver} presents the black-box solver for the feedback RT control problem while Section~\ref{sec:da_bo} formulates the outer DA profile search as a BO problem. Section~\ref{sec:adaptive_refinement} introduces the end-to-end ARBO-DART algorithmic framework.
Section~\ref{sec:numerical_results} reports the numerical results and compares the ARBO-DART algorithm with fixed-resolution baselines. Section~\ref{sec:conclusion} concludes and discusses further applications of the framework.

%%%%%%%%%%%%%%%%%%%%%%%%%%%%%
\section{DART Model and Co-optimization Objective}
\label{sec:problem_setup}

We consider a BESS participating in DA and RT electricity markets on a fixed operating day $D$. We next summarize the market setting, introduce the DA price vector and RT price process, and then define the BESS dynamics and co-optimization objective.

\subsection{U.S. Electricity Market Settlement Structure}
\label{subsec:market_design_scope}

U.S.~wholesale electricity markets generally follow a two-settlement design, where participants establish financially binding positions in the DA market; adjustments of those positions are settled in the RT market closer to physical delivery. The DA layer is hourly, whereas RT markets operate at finer temporal resolutions that varies across ISOs/RTOs. In CAISO, hourly DA profiles are followed by RT balancing in 15-minute and 5-minute markets; ancillary services (AS) are co-optimized with energy primarily in the DA market \parencite{CAISO_MarketOperationsProductsServices}. In contrast, ERCOT's RT Co-optimization plus Batteries (RTC+B) design co-optimizes energy and AS awards in each RT security-constrained economic dispatch run, approximately every five minutes, explicitly accounting for battery SoC \parencite{ERCOT_RTCB_GoLive_2025}. Other ISOs/RTOs, including PJM, NYISO, MISO, and ISONE, also combine hourly DA markets with sub-hourly RT dispatch or pricing, but differ in settlement intervals, reserve products, scarcity pricing rules, and treatment of storage resources. Nevertheless, the underlying storage control structure is common across markets: DA positions determine primary dispatch actions and available SoC headroom for BESS to respond to RT price fluctuations.

For concreteness, our numerical study uses CAISO SP-15 hub prices. The proposed co-optimization framework is not CAISO-specific: applying it to another ISO/RTO would mostly require recalibrating the DART model parameters with the mathematical foundations remaining unchanged.

%%%%%%%%%%%%%%%%%
\subsection{DA Price Vector and RT Price Process}
\label{subsec:da_vector_rt_process}

In line with the above structure of DA markets, we consider optimization over a single day, omitting inter-day dependence. Moreover, since there are often non-trivial actions in the late evening, without loss of generality we view our operational day as starting and ending at 2am. 

\noindent \textbf{DA price vector.} We denote the DA locational marginal price (LMP) vector by
\( 
\mathbf P^{\mathrm{DA}}
=
(P^{\mathrm{DA}}_0,\ldots,P^{\mathrm{DA}}_{H-1})^\top
\in \mathbb R^{H},
\)
where \(P^{\mathrm{DA}}_h\) denotes the DA price for delivery hour \(h=0,\ldots,H-1\) with \( H=24\), measured in \(\$/\mathrm{MWh}\). DA prices are commonly modeled using autoregressive specifications with calendar effects, in which each hourly price depends on recent lagged prices, day-of-week indicators, and seasonal components \parencite{lohndorf_value_2023,finnah_integrated_2022,DART_SP_Krish}. In our setting, we take $\mathbf P^{\mathrm{DA}}$ as exogenously given, acting as the input for the subsequent optimization. Thus, we assume that the BESS operator observes the DA prices and then decides on her DA charge/discharge allocations, keeping in mind headroom for RT adjustments. 

\noindent\textbf{RT price stochastic process.} While we view the DA profile as fixed, RT prices are modeled as a stochastic process that represents the intraday variability encountered during RT dispatch. For the remainder of the paper, we assume that RT prices are specified at  
15-minute resolution. Let \(\Delta t=15/60=1/4\) and \(t_k=k\Delta t\), \(k=0,\ldots,K\), with \(K=96\). We consider a filtered probability space
\( 
(\Omega,\cF,\mathbb F=(\cF_k)_{k=0,\ldots,K},\mathbb P),
\)
and model the RT price as an \(\mathbb F\)-adapted, real-valued stochastic process \(P^{\mathrm{RT}}:=(P_k^{\mathrm{RT}})_{k=0,\ldots,K}\). Thus, \(P_k^{\mathrm{RT}}\) is \(\cF_k\)-measurable for each \(k\), and the initial price \(P_0^{\mathrm{RT}}\) is \(\cF_0\)-measurable.

It is common to model RT price uncertainty using mean-reverting stochastic
dynamics, cf.~\textcite{Focker_Planck_wind_BESS,johnson_partial_2017}. 
Following \textcite{tankov_wind_BESS}, we use an additive structure for $P^{\mathrm{RT}}$, modeled as the
DA price plus a mean-zero Ornstein--Uhlenbeck (OU) factor: 
\begin{equation}
P_k^{\mathrm{RT}}
=
P^{\mathrm{DA}}_{h(k)}
+
\lambda P^{\mathrm{DA}}_{h(k)} \,Y_k,
\qquad k=0,\ldots,K-1.
\label{eq:rt_factor_price}
\end{equation}
Here \(Y_k\) is a mean-zero OU process and \( h(k) \equiv\lfloor k/4\rfloor \) denotes the DA delivery hour associated with the 15-minute RT index \(k\). In the volatility term,  \(\lambda\) controls the relative
magnitude of RT uncertainty and the scaling by
\(P^{\mathrm{DA}}_{\lfloor k/4\rfloor}\) makes RT deviations larger in absolute
terms during high-price hours. Starting with a given $Y_0$, the OU factor follows the exact discrete-time AR(1) transition law:
\begin{equation}
Y_{k+1}
=
e^{-\kappa \Delta t}Y_k
+
\sigma
\sqrt{
\frac{1-e^{-2\kappa \Delta t}}{2\kappa}
}
\,\varepsilon_{k+1}, \qquad \varepsilon_{k+1} \sim \cN(0,1)
\qquad k=0,\ldots,K-1,
\label{eq:ou_factor_dynamics}
\end{equation}
where \(\kappa>0\) is the mean-reversion rate, \(\sigma>0\) is the factor
volatility, and \((\varepsilon_k)_{k=1,\ldots,K}\) are i.i.d. standard normal
innovations.
Since \(Y_k\) has zero mean, the DART spread is mean-zero in expectation:
\(
\bE[P_k^{\mathrm{RT}}]
=
P_{h(k)}^{\mathrm{DA}}
\)
for all $k$.  

\begin{figure}[H]
    \centering
    \includegraphics[width=0.7\textwidth]{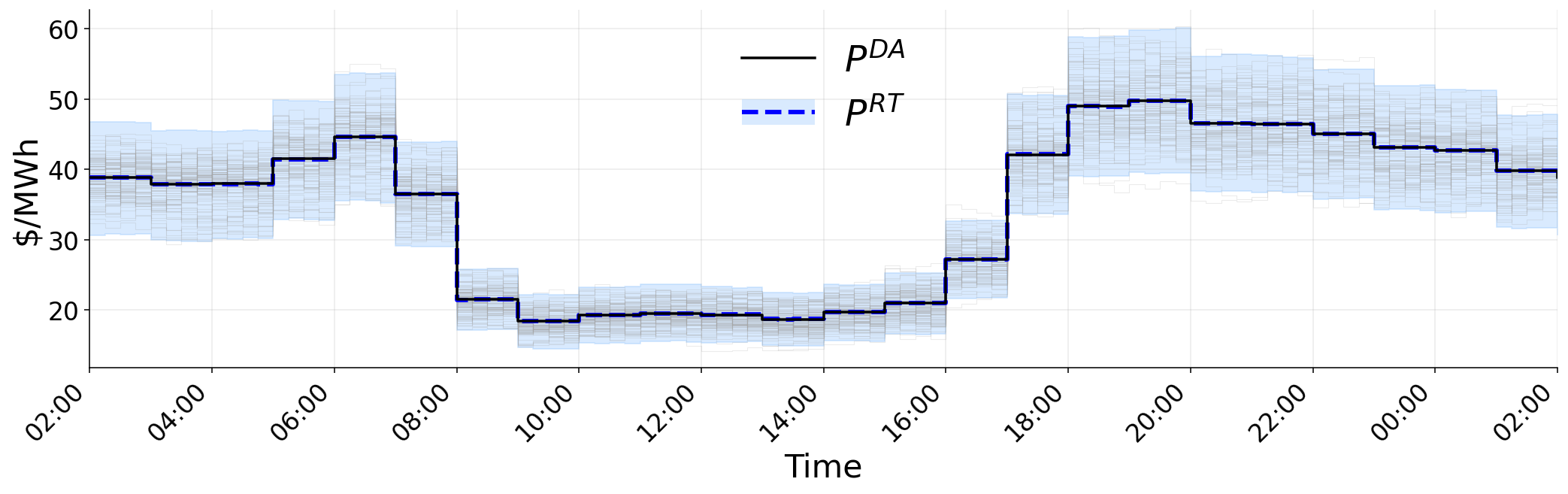}
    \caption{
RT price process $P^{\mathrm{RT}}$ at the SP-15 node for a representative day. The hourly DA prices $P^{\mathrm{DA}}_h$ (in black) coincide with the empirical average RT price path $\bE[ P^{\mathrm{RT}}_k]$ (in blue). The blue band shows the empirical $99$-th percentiles of $P^{\mathrm{RT}}$ while the gray lines are \(100\) sample paths. The RT dynamics follow \eqref{eq:rt_factor_price} with parameters \(\kappa=0.2\), \(\sigma=1.0\) and $\lambda=0.05$.
    }
    \label{fig:da_rt_price_process}
\end{figure}

Figure~\ref{fig:da_rt_price_process} illustrates our joint RT model.  The underlying DA profile $\mathbf P^{\mathrm{DA}}$ is constructed by averaging hourly historical CAISO SP-15 DA LMP prices over the 90-day window from August 3, 2025 to October 31, 2025, obtained through \textcite{gridstatus}.  The simulated RT paths of $(P^{\mathrm{RT}}_k)$ fluctuate around the DA
profile, and their mean coincides with the hourly DA
prices. The width of the $99$-th percentile band of $P^{\mathrm{RT}}_k$ varies over the day:
it is narrower during lower-price periods and wider during higher-price
periods, reflecting the multiplicative term
\(\lambda P^{\mathrm{DA}}_{h(k)}Y_k\) in \eqref{eq:rt_factor_price}.

%%%%%%%%%%%%%%%%%%%%%
\subsection{BESS Dynamics}

\paragraph{DA SoC process and feasibility.}
The DA actions are given by the hourly
power profile $\textbf{B}^{\mathrm{DA}} = (B_h^{\mathrm{DA}})_{h=0}^{H-1}$ in units MW submitted by the BESS operator. Let $(I_h^{\mathrm{DA}}(\mathbf B^{\mathrm{DA}}))_{h=0}^{H}$
denote the corresponding DA SoC trajectory with initial SoC $I_0$ in units MWh.
With $B^{\mathrm{DA}}_h>0$ denoting charging, and $B^{\mathrm{DA}}_h<0$ discharging, the SoC evolves according to:
\begin{equation}\label{eq:da_soc}
I_{h+1}^{\mathrm{DA}}(\mathbf B^{\mathrm{DA}})=I_h^{\mathrm{DA}}(\mathbf B^{\mathrm{DA}})+\eta B_h^{\mathrm{DA}} \mathbb{1}_{\left\{B_h^{\mathrm{DA}}>0\right\}}+\frac{1}{\eta} B^{\mathrm{DA}}_h \mathbb{1}_{\left\{B^{\mathrm{DA}}_h<0\right\}}.
\end{equation}
The dimensionless parameter  $\eta \in (0,1)$ denotes the charge/discharge efficiency, which means that the BESS dissipates more energy than it discharges and accumulates less energy than it charges. Typical values for $\eta$ are around 90-95\%. The BESS SoC and power constraints are given by
\begin{equation}
\label{eq:bess_bounds}
-\bar B \leq B^{\mathrm{DA}}_h \leq \bar B, \qquad 0 \leq I^{\mathrm{DA}}_h \leq I_{\max}.
\end{equation} 
Here, \(I_{\max}\) denotes the battery energy capacity in MWh, and \(\bar B>0\) denotes the symmetric charge/discharge power ratings in MW.
The corresponding feasible set of DA profiles is therefore 
\begin{equation}\label{eq:X_DA}
\cX^{\mathrm{DA}}
:=
\left\{
\mathbf B^{\mathrm{DA}}\in[-\bar B,\bar B]^{H}
:
I_h^{\mathrm{DA}}(\mathbf B^{\mathrm{DA}})\in[0,I_{\max}]\quad \forall h=0,\ldots,H
\right\}.
\end{equation}

\paragraph{RT SoC process and feasibility.}

The DA profile \(\mathbf B^{\mathrm{DA}}=(B_h^{\mathrm{DA}})_{h=0}^{H-1}\) is fixed before
RT operations begin, and hence is known at time $0$, i.e., it is \(\cF_0\)-measurable.  The BESS operator applies an adjustment process
$\Delta^{\mathrm{RT}} :=(\Delta_k^{\mathrm{RT}})_{k=0}^{K-1}$
where \(\Delta_k^{\mathrm{RT}}\) denotes the RT recourse from the DA schedule \( B_{h(k)}^{\mathrm{DA}}\) over
interval \(k\). We postulate that \(\Delta^{\mathrm{RT}}\) is an \(\mathbb F\)-adapted
closed-loop control process:
\[
\Delta_k^{\mathrm{RT}}=\pi_k(P_k^{\mathrm{RT}},I_k^{\mathrm{RT}}) \in \cF_k,
\qquad k=0,\ldots,K-1,
\]
for $\mathcal{F}_k$-measurable feedback functions \(\pi_k\). Therefore, $\Delta^{\mathrm{RT}}_k$  depends only on the current observed price $P^{\mathrm{RT}}_k$ and SoC $I^{\mathrm{RT}}_k$. Given the DA profile \(\mathbf B^{\mathrm{DA}}\) and RT adjustment process
\(\Delta^{\mathrm{RT}}\), the RT physical dispatch is
\begin{align}\label{eq:B-rt}
B_k^{\mathrm{RT}}
:=
B_{h(k)}^{\mathrm{DA}}+\Delta_k^{\mathrm{RT}},
\qquad k=0,\ldots,K-1 .
\end{align}
The corresponding RT SoC process
\(I^{\mathrm{RT}}:=(I_k^{\mathrm{RT}})_{k=0}^{K}\) then evolves according to
\begin{equation}
\label{eq:delta_rt_soc}
I_{k+1}^{\mathrm{RT}}
=
I_k^{\mathrm{RT}}
+
\Delta t
\left(
\eta B_k^{\mathrm{RT}}\mathbb{1}_{\{B_k^{\mathrm{RT}}>0\}}
+
\frac{1}{\eta}B_k^{\mathrm{RT}}\mathbb{1}_{\{B_k^{\mathrm{RT}}<0\}}
\right),
\qquad k=0,\ldots,K-1,
\end{equation}
with initial condition \(I_0^{\mathrm{RT}}=I_0\). The recourse process $(\Delta^{\mathrm{RT}}_k)_{k=0}^{K-1}$ is called admissible for the DA
profile \(\mathbf B^{\mathrm{DA}}\) if the induced dispatch and SoC satisfy the BESS
constraints almost surely:
\begin{equation}\label{eq:rt_admissibility_constraints}
-\bar B
\le
B_{h(k)}^{\mathrm{DA}}+\Delta_k^{\mathrm{RT}}
\le
\bar B,
\qquad
0\le I_k^{\mathrm{RT}}\le I_{\max},
\qquad \forall k=0,\ldots,K-1,
\quad \mathbb P\text{-a.s.}
\end{equation}
We moreover require the energy neutrality condition $I^{\mathrm{RT}}_K=I_0$ which means that the total energy charged and discharged, accounting for the efficiency losses, over the day must balance out. Note that the operator need not balance $\mathbf{B}^{DA}$, only the aggregate DART dispatch. 

Accordingly, for a DA profile $\mathbf B^{\mathrm{DA}}$, the resulting feasible set for $(\Delta_k^{\mathrm{RT}})_{k=0}^{K-1}$ is
\begin{equation}\label{eq:X_RT}
\cX^{\mathrm{RT}}(\mathbf B^{\mathrm{DA}})
:=
\left\{
(\Delta_k^{\mathrm{RT}})_{k=0}^{K-1}
\;\middle|\;
\begin{array}{l}
\Delta_k^{\mathrm{RT}}=\pi_k(P_k^{\mathrm{RT}},I_k^{\mathrm{RT}}) \text{ for } \cF_k\text{-measurable function }  \pi_k,\\[0.2em]
-\bar B \le B_{h(k)}^{\mathrm{DA}}+\Delta_k^{\mathrm{RT}}\le \bar B,\quad
0\le I_k^{\mathrm{RT}}\le I_{\max} \;\forall k, \quad I^{\mathrm{RT}}_K = I_0
\end{array}
\right\}.
\end{equation}

%%%%%%%%%%%%%%

\subsection{Optimization Objective}

The BESS operator has two sources of
profits and losses (PnL). In the DA market, the operator chooses, at $t=0$ hence deterministic, an hourly commitment vector
\(\mathbf B^{\mathrm{DA}}\) and receives the corresponding DA trading payoff. During the day, the
operator can adjust around this DA baseline in real-time in response to realized price
movements, while respecting the physical BESS constraints  encoded in the RT feasibility set \( \cX^{\mathrm{RT}}(\mathbf B^{\mathrm{DA}})\).

\paragraph{DA value.}
For a DA profile \(\mathbf B^{\mathrm{DA}}=(B_h^{\mathrm{DA}})_{h=0}^{H-1}\), the DA trading value is simply the net gain from buying and selling energy over the $H$ hours,
\begin{equation}
J^{\mathrm{DA}}(\mathbf B^{\mathrm{DA}})
=
-\sum_{h=0}^{H-1} P_h^{\mathrm{DA}} B_h^{\mathrm{DA}}.
\label{eq:j_da}
\end{equation}

Direct maximization of \eqref{eq:j_da} subject to the DA SoC dynamics
\eqref{eq:da_soc}, the energy neutrality terminal condition energy neutrality  \(I_H^{\mathrm{DA}}=I_0\),  and the feasibility set \eqref{eq:X_DA} provides a baseline for the co-optimization, in particular to judge the value of  RT recourse. This DA-only problem is a standard static optimization that can be formulated and solved as a mixed-integer linear program (MILP). Following \textcite{rolling_intrinsic_MILP}, we decompose the DA decisions into nonnegative charging and discharging variables \(c_h,d_h\ge 0\), with \(B_h^{\mathrm{DA}}=c_h-d_h\), and introduce binary variables \(z_h\in{0,1}\).
The SoC dynamics become
\begin{equation}
\label{MILP_SoC}
I_{h+1}^{\mathrm{DA}}
=
I_h^{\mathrm{DA}}
+
\Delta t
\left(
\eta c_h
-
\frac{1}{\eta}d_h
\right),
\qquad h=0,\ldots,H-1.
\end{equation}
The feasible set $\mathcal{X}^{\mathrm{DA}}$ is converted into
\begin{equation}
\label{eq:X_DA_greedy}
\cX^{\mathrm{DA}}
=
\left\{
(\mathbf c,\mathbf d,\mathbf z):
\begin{array}{l}
0 \le c_h \le \bar B z_h,\quad
0 \le d_h \le \bar B(1-z_h),\quad
z_h\in\{0,1\};\\[0.5ex]
0\le I_h^{\mathrm{DA}}\le I_{\max},\forall h=0,\ldots,H-1,\quad
I_H^{\mathrm{DA}}=I_0
\end{array}
\right\}.
\end{equation}
The DA-only policy is obtained by solving the mixed-integer linear program (MILP)
\begin{equation}
\label{eq:full_MILP_greedy}
\max_{(\mathbf c,\mathbf d,\mathbf z)\in \cX^{\mathrm{DA}}}
\quad
-\sum_{h=0}^{H-1} P_h^{\mathrm{DA}}(c_h-d_h)
\qquad
\text{subject to } \eqref{MILP_SoC},
\end{equation}
which relies on standard Branch-and-Bound methods. 

\paragraph{RT value.}
The RT recourse value associated
with an adjustment policy
\(\Delta^{\mathrm{RT}}=(\Delta_k^{\mathrm{RT}})_{k=0}^{K-1}\) is
\begin{equation}
J^{\mathrm{RT}}(\Delta^{\mathrm{RT}})
:=
\sum_{k=0}^{K-1}
\left(
- P_k^{\mathrm{RT}}\Delta_k^{\mathrm{RT}}
-\frac{\gamma}{2}(\Delta_k^{\mathrm{RT}})^2
\right)\Delta t
-\frac{\rho}{2}\bigl(I_{K}^{\mathrm{RT}}-I_{0}\bigr)^2.
\label{eq:j_rt}
\end{equation}
The first term represents the RT trading payoff from incremental deviations around the DA baseline. Next, the running cost parameter \(\gamma\) penalizes aggressive RT adjustments and can be interpreted as a reduced-form trading friction or price-impact cost. Larger values of \(\gamma\)  impose stronger regularization on the RT policy and shift more dispatch responsibility to the DA profile. The penalty parameter \(\rho\) enforces the energy neutrality SoC requirement by penalizing deviations of the terminal SoC from \(I_0\).  We 
tend to use a large terminal penalty, \(\rho=100\), which keeps \(I_K^{\mathrm{RT}}\) very close to the initial state of charge across all case studies. We emphasize that the linear-quadratic form in \eqref{eq:j_rt} is used for convenience and is not
essential to the proposed framework. Other specifications such as non-quadratic trading
frictions can be incorporated as long as the resulting RT control problem can be solved by the inner policy solver.

\paragraph{Coupled DART value.}
Combining the DA trading value and the RT recourse value, the DART co-optimization
problem is
\begin{equation}
\label{eq:dart_exact}
\sup_{\mathbf B^{\mathrm{DA}}\in\cX^{\mathrm{DA}}}
\left\{
J^{\mathrm{DA}}(\mathbf B^{\mathrm{DA}})
+
\sup_{\Delta^{\mathrm{RT}}\in\cX^{\mathrm{RT}}(\mathbf B^{\mathrm{DA}})}
\mathbb E\!\left[
J^{\mathrm{RT}}(\Delta^{\mathrm{RT}})
\right]
\right\}.
\end{equation}
The outer optimization is carried out over the static feasible set
$\cX^{\mathrm{DA}}$ of DA commitment vectors in \eqref{eq:X_DA}, while the inner optimization is taken over the dynamic and $\mathbf B^{\mathrm{DA}}$-dependent set
$\cX^{\mathrm{RT}}(\mathbf B^{\mathrm{DA}})$  of admissible RT feedback control policies in \eqref{eq:X_RT}. 

The formulation \eqref{eq:dart_exact} is bilevel in structure, but it is not a standard
finite-scenario two-stage program: the outer problem is a finite-dimensional
optimization over DA profiles, whereas the inner problem is an optimization over
the space of closed-loop RT policy maps. In particular, the two levels are coupled through the DA profile $\mathbf B^{\mathrm{DA}}$,
which enters the RT dispatch and the induced RT SoC evolution and RT constraints \eqref{eq:delta_rt_soc}-\eqref{eq:rt_admissibility_constraints}. The DA
problem is governed by the DA SoC process $I^{\mathrm{DA}}$ \eqref{eq:da_soc} and DA price vector  $\mathbf{P}^{\mathrm{DA}}$,
whereas the RT problem is driven by the RT SoC process $I^{\mathrm{RT}}$
\eqref{eq:delta_rt_soc} and the RT price process $P^{\mathrm{RT}}$ 
\eqref{eq:rt_factor_price}.

\paragraph{Total trading profit.}
We define the running DART PnL, excluding the running
friction cost and terminal SoC penalty from RT objective \eqref{eq:j_rt} , as
\begin{equation}
\Pi_k^{\mathrm{DART}}
:=
-\sum_{h=0}^{\lfloor k/4\rfloor} P_h^{\mathrm{DA}} B_h^{\mathrm{DA}}
-\sum_{j=0}^{k} P_j^{\mathrm{RT}}\Delta_j^{\mathrm{RT}}\Delta t,
\qquad k=0,\ldots,K-1.
\label{eq:pi_dart_explicit}
\end{equation}
Thus, \(\Pi_{K-1}^{\mathrm{DART}}\) is the DART PnL over the full horizon. For comparison, we define the running DA PnL as
\begin{equation}
\Pi_h^{\mathrm{DA}}
:=
-\sum_{j=0}^{h} P_j^{\mathrm{DA}}B_j^{\mathrm{DA}},
\qquad h=0,\ldots,H-1.
\label{eq:pi_da_only}
\end{equation}
with \(\Pi_{H-1}^{\mathrm{DA}}\) the DA PnL over the full day.

\section{RT Recourse Solver Setup}
\label{sec:rt_solver}
For a fixed DA profile \(\mathbf B^{\mathrm{DA}}\), we define the RT recourse value function:
\begin{equation}
V^{\mathrm{RT}}(P^{\mathrm{RT}}_0,I_0; {\mathbf B^{\mathrm{DA}}})
:=
\sup_{\Delta^{\mathrm{RT}}\in\cX^{\mathrm{RT}}(\mathbf B^{\mathrm{DA}})}
\mathbb E\!\left[
J^{\mathrm{RT}}(\Delta^{\mathrm{RT}})
\right].
\label{eq:rt_value_function}
\end{equation}
The RT problem \eqref{eq:rt_value_function} does not admit a closed-form solution due to the piecewise-linear BESS dynamics and state-dependent SoC constraints, which make the admissible RT control set depend on the current state and on $\mathbf{B}^{\mathrm{DA}}$. As a result, 
the optimal feedback policy and $V^{\mathrm{RT}}$ must be approximated numerically. This can be done using function approximation methods such as reinforcement learning (RL) or regression Monte Carlo (RMC). In this work, we use an adaptation of the actor-critic RMC SHADOw-GP algorithm developed in our previous work \parencite{Aung_Ludkovski_CDC}.  SHADOw-GP is based on the Bellman recursion implied by the dynamic programming principle (DPP) 
\begin{multline}
\label{eq:Bellman_DP}
V_k^{\mathrm{RT}}(P,I; \mathbf B^{\mathrm{DA}})
=
\max_{\Delta_k^{\mathrm{RT}}\in \cX^{\mathrm{RT}}(B_{h(k)}^{\mathrm{DA}},I)}
\Bigg\{
\left(
- P\cdot \Delta_k^{\mathrm{RT}}
-\frac{\gamma}{2}(\Delta_k^{\mathrm{RT}})^2
\right)\Delta t \\
+
\mathbb E\!\left[
V_{k+1}^{\mathrm{RT}}(P_{k+1}^{\mathrm{RT}},I_{k+1}^{\mathrm{RT}}; \mathbf B^{\mathrm{DA}})
\,\middle|\, P_k^{\mathrm{RT}}=P, I_k^{\mathrm{RT}} = I
\right]
\Bigg\},
\qquad k=0,\ldots,K-1,
\end{multline}
with terminal condition
\( V_K^{\mathrm{RT}}(P,I; \mathbf B^{\mathrm{DA}})
=
-\frac{\rho}{2}\bigl(I-I_{0}\bigr)^2\).
The expectation in \eqref{eq:Bellman_DP} is taken over the next-step RT price
\(P_{k+1}^{\mathrm{RT}}\), conditional on the current \(P_k^{\mathrm{RT}}\). The one-step feasible set in \eqref{eq:Bellman_DP} is the pointwise
control constraint induced by the global admissibility condition
\eqref{eq:rt_admissibility_constraints}:
\begin{equation}
\label{eq:x_k}
\cX^{\mathrm{RT}}(B_{h(k)}^{\mathrm{DA}},I)
:=
\left\{
\Delta_k \in\mathbb R:
\max\left(
\bar{B},
\frac{-\eta I}{\Delta t}
\right)
\le
B_{h(k)}^{\mathrm{DA}}+\Delta_k
\le
\min\left(
\bar{B},
\frac {\eta^{-1}(I_{\max}-I)}{\Delta t}
\right)
\right\}.
\end{equation}
In line with the DPP approach, we 
define the cost-to-go $q$-value
\begin{equation}\label{eq:Q}
    Q_k(P^{\mathrm{RT}}_k,I^{\mathrm{RT}}_{k+1}):= \EE \Bigl[V^{\mathrm{RT}}_{k+1}( P^{\mathrm{RT}}_{k+1}, I^{\mathrm{RT}}_{k+1}; {\mathbf B^{\mathrm{DA}}}) \, \Big| \, P^{\mathrm{RT}}_k\Bigr]
\end{equation}
and characterize the optimal control at step $k$ via the feedback form 
\begin{multline}\label{eq:optimal}
\pi_k^*(P, I)
:=
\argmin_{\Delta_k\in \cX^{\mathrm{RT}}(B_{h(k)}^{\mathrm{DA}}, I)}
\Bigl\{
(-P \cdot \Delta_k
-\tfrac{\gamma}{2}(\Delta_k)^2)\Delta t
\\ 
+ 
Q_k\Bigl( P,
  I
+ \bigl(B_{h(k)}^{\mathrm{DA}}+\Delta_k\bigr)(\eta
\mathbf 1_{\{B_{h(k)}^{\mathrm{DA}}+\Delta_k>0\}}
+\frac{1}{\eta}
\mathbf 1_{\{B_{h(k)}^{\mathrm{DA}}+\Delta_k<0\}} ) \Delta t
\Bigr)
\Bigr\}.
\end{multline}

The algorithm approximates both the cost-to-go function \(Q_k\) and the optimal feedback policy \(\pi_k^*\) at each time step using two Gaussian-process emulators in an actor--critic fashion, denoted by \(\widehat Q_k(\cdot,\cdot)\) and \(\widehat{\pi}_k(\cdot,\cdot)\), respectively. 
The emulators are trained at each time step using step-dependent training domains that reflect the range of states reachable at that stage of the backward recursion. The training procedure is governed by three hyperparameters. For each continuation-value emulator \(\widehat Q_k\), we use \(\mathcal{N}_{\mathrm{loc}}\) Latin hypercube sampling (LHS) design points per time step and \(\mathcal{N}_{\mathrm{rep}}\) Monte Carlo replications at each design point for variance reduction. For each policy emulator \(\widehat{\pi}_k\), we use \(\mathcal{N}_{b}\) LHS design points. The LHS designs provide space-filling coverage of the corresponding state domains and reduce clustering relative to uniform random sampling, thereby improving the statistical efficiency of GP training. Further details 
are in \textcite{aung2025intradaybatterydispatchhybrid}.

\textbf{Estimating the value function of RT recourse. }
After training, SHADOw-GP produces a collection of continuation-value
emulators \(\{\widehat Q_k(\cdot,\cdot,\cdot)\}_{k=0}^{K-1}\) and control
emulators \(\{\widehat \pi_k  (\cdot,\cdot)\}_{k=0}^{K-1}\). To evaluate the
resulting policy and the associated value function, we perform out-of-sample
Monte Carlo simulation. Given an initial state \((P_0^{\mathrm{RT}}, I_0^{\mathrm{RT}})\) and 
\(\mathbf B^{\mathrm{DA}}\), we simulate \(M\) independent price paths
\(( P_k^{RT,m})_{k=0}^{K-1}\), \(m=1,\ldots,M\), according to the RT price
dynamics \eqref{eq:rt_factor_price}. Along each path, the RT adjustment is computed from the learned
feedback policy as
\begin{equation}
\label{eq:deltahat_mc}
\widehat{\Delta}_k^{\,m}
:=
\widehat \pi_k\!\left(P_k^{RT,m}, I_k^{RT,m}\right),
\qquad k=0,\ldots,K-1.
\end{equation}
The SoC is then updated recursively using the RT SoC dynamics
\eqref{eq:delta_rt_soc}, yielding $M$ independent SoC paths \( (\widehat{I}^{RT,m}_k)_{k=0}^{K-1} \). This yields the pathwise realized RT payoff
\begin{equation}
\label{eq:pathwise_rt}
\widehat v^{\,m}_{RT}
=
\sum_{k=0}^{K-1}
\left(
- P_k^{RT,m}\widehat{\Delta}_k^{\,m}
-\frac{\gamma}{2}(\widehat{\Delta}_k^{\,m})^2
\right)\Delta t
-
\frac{\rho}{2}\bigl(\widehat{I}_{K}^{RT,m}-I_{0}\bigr)^2.
\end{equation}
We obtain the Monte Carlo estimator of the RT value
\eqref{eq:rt_value_function} as the empirical average
\begin{equation}
\label{eq:value_mc_rt}
\widehat V^{\mathrm{RT}}( P^{\mathrm{RT}}_0, I^{\mathrm{RT}}_0; {\mathbf B^{\mathrm{DA}}})
=
\frac{1}{M}\sum_{m=1}^{M}\widehat v^{\,m}_{RT}.
\end{equation}
 Unless otherwise noted, the RT value $\widehat{V}^{\mathrm{RT}}$ is estimated using
\(M=10{,}000\) Monte Carlo paths for evaluation. 

%%%%%%%%%%%%
\section{ARBO-DART}
\label{sec:da_bo}
\subsection{DA-Only Solution Structure}\label{sec:da-only}
\begin{figure}[t]
\centering
\includegraphics[width=0.8\linewidth]{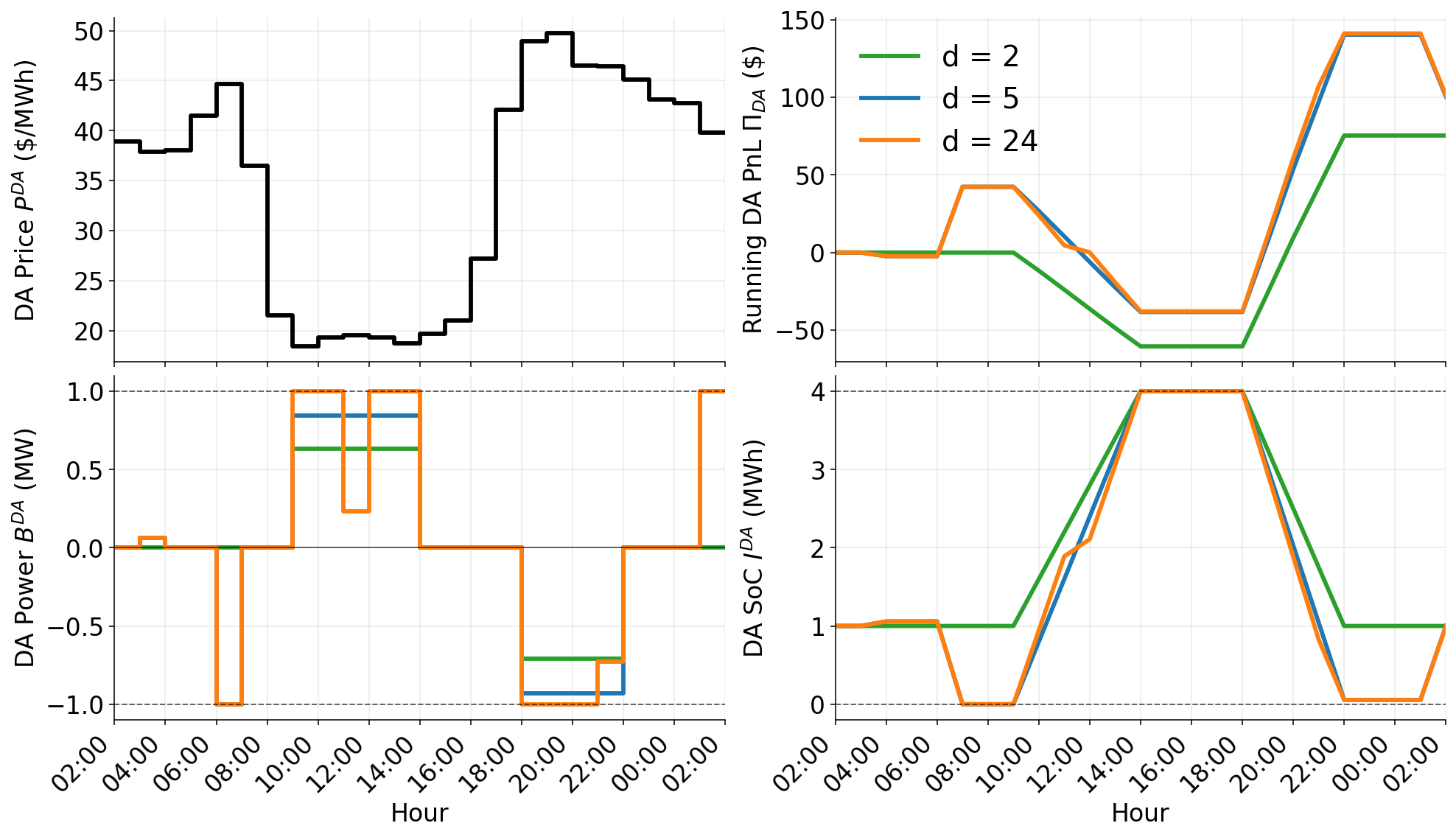}
\caption{Optimal DA dispatch under three parameterizations using $d=2,5$ and the full $d=24$ blocks.  Model parameters are $I_{\max}=4$MWh, $\bar{B}=1$MW, $\eta=0.95$, initial and terminal SoC $I_H=I_0=1$MWh. 
\emph{Top left}: DA price curve $(P^{\mathrm{DA}}_h)$. \emph{Top right}: running DA PnL $(\Pi^{\mathrm{DA}}_h)$. \emph{Bottom left}: DA charge/discharge profile $(B^{\mathrm{DA}}_h)$. \emph{Bottom right}: DA SOC trajectory $(I^{\mathrm{DA}}_h)$. The dashed horizontal lines indicate maximum charge/discharge limits $\pm\bar{B}$ and energy capacity $I_{\max}$. }
\label{fig:DA_greedy_comparator}
\end{figure}

We now consider the outer problem of optimizing the $H$-dimensional DA dispatch commitment \(\mathbf{B}^{\mathrm{DA}}\in\cX^{\mathrm{DA}}\).
Since the RT recourse value is evaluated numerically, we rely on the 
Monte Carlo estimator \(\widehat V^{\mathrm{RT}}(\mathbf B^{\mathrm{DA}})\) from \eqref{eq:value_mc_rt}. This gives the empirical version of  DART objective, denoted by
\begin{equation}\label{eq:hat-J}
    \widehat\cJ(\mathbf{B}^{\mathrm{DA}}) := J^{\mathrm{DA}}(\mathbf{B}^{\mathrm{DA}}) + \widehat{V}^{\mathrm{RT}}(\mathbf{B}^{\mathrm{DA}})
\end{equation} 
which is 
\begin{itemize}
    \item expensive to evaluate, since each evaluation of $\mathbf{B}^{\mathrm{DA}}$ requires solving the inner RT problem;
    \item statistically noisy, due to the Monte Carlo errors in $\widehat{V}^{\mathrm{RT}}(\mathbf{B}^\mathrm{DA})$ and the approximation errors stemming from
    $\{\widehat{Q}_k(\cdot,\cdot),\widehat{\pi}_k(\cdot,\cdot)\}_{k=0}^{K-1}$;
    \item non-convex since the RT recourse value depends on $\mathbf{B}^{\mathrm{DA}}$ through a nonlinear constrained stochastic control problem.
\end{itemize}
To address these challenges that  make classical optimization methods ill-suited for our formulation, we develop a novel solution approach that relies on BO. Our approach is motivated by the solution structure of the DA-only baseline \eqref{eq:full_MILP_greedy}, solved via MILP, which exhibits a bang--bang pattern with nonzero DA allocations organized into a small number of charging and discharging blocks, see bottom left panel of Figure~\ref{fig:DA_greedy_comparator} for $d=24$. The round-trip efficiency loss imposes friction, so a charge/discharge cycle is only undertaken when the price spread yields positive profit after accounting for $\eta$-cost. 
Furthermore, the resulting hourly solution has just eight nonzero charge/discharge blocks, suggesting that a lower-dimensional blockwise parametrization can capture much of the profit-generating structure of the full MILP solution. 

We re-solve the DA-only MILP under a \(2\)-block $(d=2)$ parameterization, restricting nonzero allocations to 09:00–14:00 and 18:00–22:00. This choice is motivated by the DA price profile in the top-left panel of Figure~\ref{fig:DA_greedy_comparator}: the battery charges during the low-price window and discharges during the high-price window, similar to the TB4 benchmark. Here, TB-{k} denotes a top-bottom benchmark that charges maximally during the $k$ lowest-price blocks and discharges maximally during the $k$ highest-price blocks, assuming \(k<I_{\max}/\bar{B}\). The \(2\)-block parameterization yields a daily PnL of \(\$76.40\), compared with \(\$76.77\) for TB4 and \(\$101.78\) for the fully flexible hourly solution. The performance loss relative to the hourly solution arises from the restricted profile, which misses both an earlier charging opportunity around 06:00 and low-price charging near the end of the day needed to return the battery to its terminal SoC target \(I_0\).
 
Figure~\ref{fig:DA_greedy_comparator} also shows a \(5\)-block parameterization $(d=5)$ solution that refines the \(2\)-block structure by adding three additional single-hour actions at hours 1:00, 3:00, and 6:00, in addition to the two multi-hour windows, 9:00--14:00 and 18:00--22:00. 
As shown in top right panel, this refinement increases the DA PnL from \(\$76.40\) to \(\$101.10\) which is less than 70 cents away from the hourly-resolution solution. The comparison among the \(2\)-, \(5\)-, and \(24\)-block DA MILP solutions suggests that, for the DA-only problem, a piecewise-constant profile with relatively few intervals can closely approximate the fully flexible hourly solution, provided that the main charging and discharging intervals are identified. This motivates us to apply the same low-dimensional blockwise parameterization for DART co-optimization, forming the basis of our algorithm described in the next section.

%%%%%%%%%%%%
\subsection{Piecewise Constant DA Formulation}
To reduce the dimension of the DA search space, we represent the \(H\)-hour
horizon by a signed partition
\( 
\cP :=
\left\{
[a_j,b_j)_{s_j}
:\;
0 \le a_j < b_j \le H,\;
s_j \in \{+1,-1\}
\right\}_{j=1}^d.
\)
where \(s_j=+1\) denotes charging and \(s_j=-1\) denotes discharging. The
endpoints are indexed relative to the model horizon, with index \(0\)
corresponding to \(02{:}00\) in Figure~\ref{fig:DA_greedy_comparator} and all
subsequent figures.

Each block \(j\) has length \(L_j=b_j-a_j\) and constant DA power
\(s_jx_j\), where \(x_j\ge 0\). The block amplitudes therefore define the
\(d\)-dimensional DA decision vector
\(
\bx=(x_1,\ldots,x_d)\in\mathbb{R}_+^d.
\)
For notational convenience, let
\( 
w_j:=\eta^{s_j},
\)
so that \(w_j=\eta\) for charging blocks and \(w_j=1/\eta\) for discharging
blocks. The energy and power constraints then define the feasible set of DA
profiles over \(\cP\) as the convex polytope:
\[
\mathcal{X}^{\mathrm{DA}}(\cP)
=
\Big\{
\bx \in \mathbb{R}_+^d :
0 \le I_0 + \sum_{j=1}^{q} s_j L_j w_j x_j \le I_{\max},
\ \;
0 \le x_j \le \bar{B}, \ \forall   q =1,\ldots,d
\Big\},
\]
where $I_0 + \sum_{j=1}^{q} s_j L_j w_j x_j$ is the SoC at the end of the $q$-th block. Thanks to the pre-assigned signs \(s_j\), the SoC constraints become linear rather than piecewise linear. Consequently, for a fixed partition \(\cP\), the DA problem can be solved without the integer variables $\bz$ required in the MILP benchmark \eqref{eq:full_MILP_greedy}. We therefore may consider
\begin{equation}
\bx^{*}(\cP)
\in
\argmax_{\bx\in\mathcal{X}^{\mathrm{DA}}(\cP)}
\widehat{\mathcal{J}}(\bx),
\label{eq:outer_problem_fixed_partition}
\end{equation}
where \(\bx\) is a \(d\)-dimensional decision variable with \(d\le H\), and \(\mathcal{X}^{\mathrm{DA}}(\cP)\) encodes the corresponding linear SoC constraints.
However, \eqref{eq:outer_problem_fixed_partition} is still a statistically noisy, expensive to evaluate, non-convex objective. This motivates the use of BO, which (i) constructs a
probabilistic cheap-to-evaluate surrogate of the objective function, and (ii) chooses the next evaluation candidate via a so-called acquisition function which balances exploration and exploitation. 
In the next section, we discuss Gaussian process regression and 
the Upper Confidence Bound acquisition function utilized for these two respective sub-steps.

\subsection{Gaussian Process Regression}

Given a fixed partition \(\cP\), we place a Gaussian process (GP) prior on the objective function \(\cJ:\cX^{\mathrm{DA}}(\cP)\to\mathbb R\) in \eqref{eq:hat-J}. Hence, for any finite collection of inputs \(\bx^{(1)},\ldots,\bx^{(n)}\in \cX^{\mathrm{DA}}(\cP)\), the random vector \((\cJ(\bx^{(1)}),\ldots,\cJ(\bx^{(n)}))\) is assumed to be jointly Gaussian. After \(n\) evaluations, we have the dataset \(\cD_n=\{(\bx^{(i)},y^{(i)})\}_{i=1}^n\), where \(y^{(i)}:=\widehat{\cJ}(\bx^{(i)})\). We assume that estimated $\widehat{V}^{\mathrm{RT}}$ values lead to
\(y^{(i)}=\cJ(\bx^{(i)})+\varepsilon_i\), with \(\varepsilon_i\sim\cN(0,\sigma_\varepsilon^2)\) independently across \(i\), where \(\sigma_\varepsilon^2\) denotes the observation noise variance arising from Monte Carlo estimation of the RT-stage value \eqref{eq:value_mc_rt}. In our numerical implementation, $\sigma_{\varepsilon}^2$ is estimated from the RT-stage simulations and then treated as fixed throughout the GP fitting subroutine. The GP prior is specified by a mean function
$m(\textbf{x})$ and covariance function $c(\textbf{x},\textbf{x'}; \vartheta)$ where hyperparameters $\vartheta$ of the covariance kernel specify the smoothness of $\mathcal{{J}}(\cdot)$. We use the anisotropic 
Mat\'ern-\(5/2\) covariance kernel, default kernel choice for BO. Furthermore, this twice differentiable kernel aligns with our quadratic objective \eqref{eq:dart_exact}. For
\(\bx,\bx'\in\mathcal{X}^{\mathrm{DA}}(\cP)\), this kernel is given by
\[
c_{M52}(\bx,\bx';\vartheta)
=
\sigma_p^2
\left(
1+\sqrt{5}\,r(\bx,\bx')
+\frac{5}{3}r(\bx,\bx')^2
\right)
\exp\!\bigl(-\sqrt{5}\,r(\bx,\bx')\bigr),
\]
where
\[
r(\bx,\bx')
=
\left(
\sum_{j=1}^d \frac{(x_j-x_j')^2}{\ell_j^2}
\right)^{1/2},
\]
and \(\vartheta=(\sigma_p^2,\ell_1,\ldots,\ell_d)\) denotes the kernel
hyperparameters, with \(\sigma_p^2\) the prior signal variance and
\(\ell_1,\ldots,\ell_d\) the coordinate-wise lengthscales. Conditioning on the data \(\cD_n\), the GP posterior at a
 \(\bx_*\in\mathcal{X}^{\mathrm{DA}}(\cP)\) is Gaussian with posterior mean and variance
\[
\mu_n(\bx_*)
=
m(\bx_*)
+
\bc_*^\top
\bigl(\bC+\sigma_\varepsilon^2 \bI\bigr)^{-1}\bz,
\qquad
\sigma_n^2(\bx_*)
=
c(\bx_*,\bx_*)
-
\bc_*^\top
\bigl(\bC+\sigma_\varepsilon^2 \bI\bigr)^{-1}\bc_*.
\]
where $
\bz
=
\bigl(y^{(1)}-m(\bx^{(1)}),\ldots,y^{(n)}-m(\bx^{(n)})\bigr)^\top
$, 
$
\bc_*
=
\bigl(
c(\bx_*,\bx^{(1)}),\ldots,
c(\bx_*,\bx^{(n)})
\bigr)^\top
$,
$\bC \in \mathbb{{R}}^{n\times n}$ is the covariance matrix with entries
$C_{ij}=c_{M52}(\bx^{(i)},\bx^{(j)}),
 i,j=1,\ldots,n,
$ and $\bI$ is the $n\times n$ identity matrix. The hyperparameter vector $\vartheta$ is optimized using maximum likelihood estimation (MLE). The posterior mean \(\mu_n(\bx_*) = \mathbb{E}[ \cJ(\bx_*) | \cD_n]\) provides a prediction of the
objective value at \(\bx_*\), while the posterior variance
\(\sigma_n^2(\bx_*)\) quantifies the uncertainty of that prediction.
In the next section, we show how the posterior mean and variance are used to construct the UCB acquisition function.

\subsection{Bayesian Optimization for DA Profiles}

\paragraph{Initial training design.} To initialize the GP surrogate for BO, we first generate and
evaluate \(n_0\) initial DA profiles obtained
 via Latin Hypercube Sampling (LHS) with a rejection scheme to target the polytope \(\mathcal{X}^{\mathrm{DA}}(\cP)\). For each sampled profile \(\bx^{(i)}, i=1,\ldots,n_0\), we compute \(y^{(i)}=\widehat{\cJ}(\bx^{(i)})\) using SHADOw-GP, forming initial dataset
\(
\cD_{n_0}
=
\{(\bx^{(i)},y^{(i)})\}_{i=1}^{n_0}.
\)

\paragraph{UCB acquisition rule.}
We impose a budget of at most $n_{max}$ evaluations on the RT solver. At each BO iteration \(n=n_0,\ldots,n_{\max}-1\), given the current dataset \(\cD_n=\{(\bx^{(i)},y^{(i)})\}_{i=1}^{n}\), we fit a GP surrogate for the objective, yielding posterior mean \(\mu_n(\bx)\) and posterior standard deviation \(\sigma_n(\bx)\). We then define the upper confidence bound (UCB) acquisition function by
\begin{equation}
\mathrm{ucb}_n(\bx)
:=
\mu_n(\bx)
+
\sqrt{\beta_n}\,\sigma_n(\bx),
\label{eq:ucb_final}
\end{equation}
where \(\beta_n>0\) is an iteration-dependent parameter controlling the trade-off between exploration and exploitation. The posterior mean \(\mu_n(\bx)\) favors points predicted to have high objective value under the current GP surrogate, while the posterior standard deviation \(\sigma_n(\bx)\) favors points in regions where uncertainty remains large. Larger values of \(\beta_n\) place more weight on exploration, whereas smaller values emphasize exploitation. The next DA candidate profile is selected at BO iteration \(n\) by maximizing the acquisition function over the feasible set,
\begin{equation}
\bx^{(n+1)}
\in
\argmax_{\bx\in\mathcal{X}^{\mathrm{DA}}(\cP)}
\mathrm{ucb}_n(\bx),
\label{eq:BO_opt}
\end{equation}
after which we evaluate \(y^{(n+1)}=\widehat{\cJ}(\bx^{(n+1)})\) and update the dataset, yielding \(\cD_{n+1}\).
In contrast to direct optimization of \eqref{eq:outer_problem_fixed_partition}, which entails repeated calls to the expensive objective $\widehat{\mathcal{J}}(\cdot)$, maximizing \(\mathrm{ucb}_n(\cdot)\) is computationally inexpensive since it is evaluated from the GP surrogate. The latter optimization is carried out in BoTorch \textcite{botorch} using \texttt{optimize\_acqf}, which by default relies on SciPy's SLSQP solver.

\paragraph{Termination criterion.}
In addition to the $n_{max}$ evaluation budget, we impose an early termination rule. To this end, define the lower confidence bound by \(\mathrm{lcb}_n(\bx):=\mu_n(\bx)-\sqrt{\beta_n}\,\sigma_n(\bx)\), and set
\begin{equation}
\bar r_n
:=
\sup_{\bx\in\cX^{\mathrm{DA}}(\cP)} \mathrm{ucb}_n(\bx)
-
\max_{\bx'\in \cD_n} \mathrm{lcb}_n(\bx').
\label{eq:regret_upper_bound_max}
\end{equation}
The regret \(\bar r_n\) is a computable upper bound on the simple regret at iteration \(n\), in the same spirit as in \textcite{BO_auto_rule}. Indeed, the first term is our UCB solution \eqref{eq:ucb_final}, while the second term, \(\max_{\bx\in \cD_n} \mathrm{lcb}_n(\bx)\), is a lower confidence bound on the best objective value already identified among the evaluated points. Consequently, \(\bar r_n\) upper bounds the maximum plausible improvement still remaining, and BO is terminated once this bound falls below a prescribed tolerance \(\varepsilon_{\mathrm{tol}}>0\) or once the evaluation budget \(n_{\max}\) is exhausted.  In this paper, we set
\(
\beta_n
=
\frac{2}{5}\log\!\left(\frac{d\,n^2\pi^2}{6\delta}\right),
\)
with \(\delta=0.1\), where \(d\) denotes the dimension of the decision vector \(\bx\in\mathcal{X}^{\mathrm{DA}}(\cP)\) as in \textcite{BO_Prop_Regret_Boundds}. At termination, after doing \(n_{\mathrm{term}}\)  evaluations of the RT recourse, we select the final DA profile:
\begin{equation}
\bx^*(\cP)
=
\argmax_{\bx^{(i)}\in\cD_{n_{\mathrm{term}}}}
y^{(i)}
\label{eq:bo_final_recommendation}
\end{equation}
yielding the corresponding estimated objective value as
\( 
y^*(\cP)\) and the associated RT policy
\(\{\hat{\pi}_k^{\bx^*(\cP)}(\cdot,\cdot)\}_{k=0}^{K-1}\).
The full BO procedure over a fixed partition is summarized in
Algorithm~\ref{alg:bo_fixed_partition}.

\begin{figure}[!htb]
    \centering
    \includegraphics[width=0.9\linewidth]{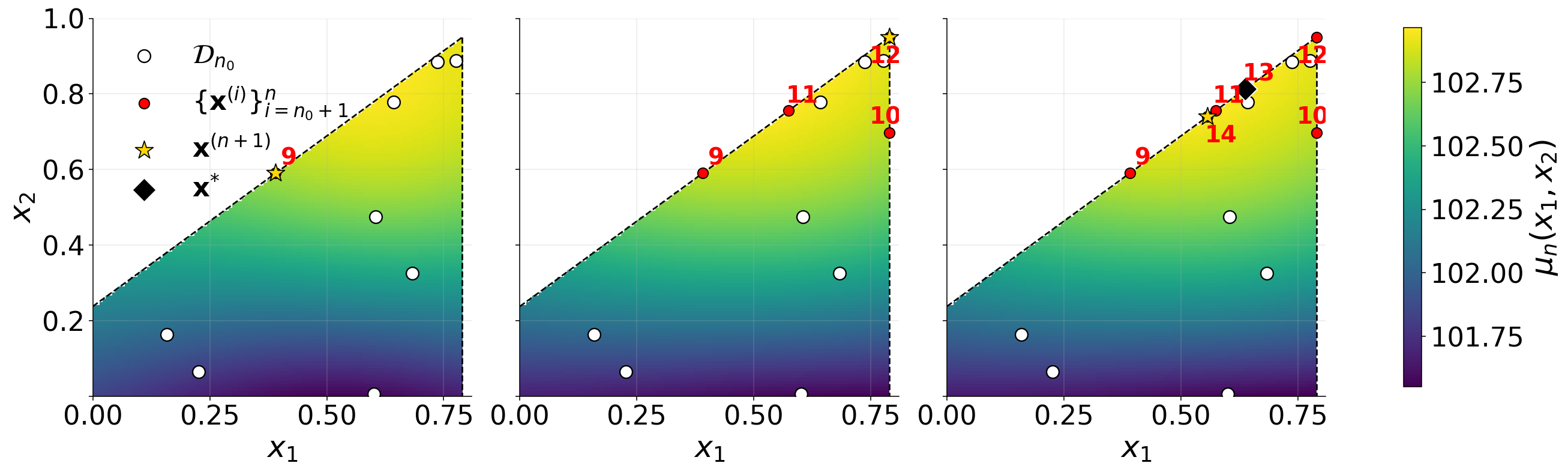}
    \caption{Illustrating DART co-optimization over 2-block parameterization $d=2$. The shading shows the GP posterior mean surfaces \(\mu_n(x_1,x_2)\) over the feasible polytope \(\cX^{\mathrm{DA}}(\cP)\) at representative BO iterations $n=8, 11, 13$. White circles denote initial LHS design with $n_0 = 8$, red circles denote $\bx^{(i)}, i=n_0+1,\ldots$, the star denotes the newly proposed candidate $\bx^{(n+1)}$ and the diamond denotes the ultimate $\bx^\star$.
    \label{fig:bo_mu_panels}}
\end{figure}

We illustrate a run of the BO algorithm in Figure~\ref{fig:bo_mu_panels}.
We consider the partition
\(
\cP=\{[9,13)_{+},[16,20)_{-}\},
\)
chosen in the spirit of TB4 (charging during the lowest 4-hour block, discharging during the highest 4-hour block), with the two decision variables \(x_1\) which is the charging rate during 11am:00-3:00pm, and \(x_2\)
which is the magnitude of the discharging rate during 6:00-10:00pm. With the battery configuration of \(I_0=1\),
\(I_{\max}=4\), \(\eta=0.95\), and \(\bar B=1\), the feasible polytope of DA actions is (cf. the trapezoid domain in Figure \ref{fig:bo_mu_panels})
\begin{align*}
\mathcal{X}^{\mathrm{DA}}(\cP)
& =
\left\{
(x_1,x_2) :
0\le 1+4(0.95)x_1\le 4,0\le 1+4(0.95)x_1-\frac{4}{0.95}x_2\le 4,0\le x_1\le 1, 0\le x_2\le 1
\right\} \\
& =
\left\{
(x_1,x_2)\in\mathbb{R}_+^2:
0\le x_1\le 0.7895,
\;
0\le x_2\le 0.2375+0.9025x_1
\right\}.
\end{align*}
The BO procedure is initialized with \(n_0=8\) initial design points in the polytope \(\cX^{\mathrm{DA}}(\cP)\), and is then run with maximum budget \(n_{\max}=38\) and stopping tolerance \(\varepsilon_{\mathrm{tol}}=1\times10^{-1}\).

\begin{algorithm}[H]
\caption{BO over a fixed partition \(\cP\)}
\label{alg:bo_fixed_partition}
\begin{algorithmic}[1]
\Require \(\mathcal{X}^{\mathrm{DA}}(\cP)\), \(n_0\), \(n_{\max}\), \(\varepsilon_{\mathrm{tol}}\)
\State Set $n=n_0$, generate \(n_0\) LHS design points in \(\mathcal{X}^{\mathrm{DA}}(\cP)\), and form
\(
\cD_{n_0}
=
\{(\bx^{(i)},y^{(i)})\}_{i=1}^{n_0}
\)
\While{\(n < n_{\max}\) and $\bar r_n > \varepsilon_{\mathrm{tol}}$}
    \State Fit a GP surrogate on \(\cD_n\)
    \State Compute UCB maximizer \(\bx^{(n+1)}\) as in \eqref{eq:BO_opt} and evaluate \(y^{(n+1)}=\widehat{\cJ}(\bx^{(n+1)})\)
 
    \State Update
    \(
    \cD_{n+1}
    =
    \cD_n
    \cup
    \left\{
    \bigl(\bx^{(n+1)},y^{(n+1)}\bigr)
    \right\}
    \)
     \State Compute regret \(\bar r_n\) as in \eqref{eq:regret_upper_bound_max}
\EndWhile
\State Set \(n_{\mathrm{term}}=n+1\) and fit the final GP surrogate on \(\cD_{n_{\mathrm{term}}}\)
\State \Return \(\bigl({\textbf{x}^*}(\cP),\,{y}^*(\cP)\bigr)\) as in \eqref{eq:bo_final_recommendation} and the corresponding RT policy \(\{\hat{\pi}_k^{{\textbf{x}^*}(\cP)}(\cdot,\cdot)\}_{k=0}^{K-1}\) 
\end{algorithmic}
\end{algorithm}

Figure~\ref{fig:bo_mu_panels} displays the posterior mean surfaces \(\mu_n\) over the feasible polytope at representative early, intermediate, and late stages of the BO procedure, namely ($n=8,11,13$). In the left panel, \(\mu_8(x_1, x_2)\) is fitted using only the 8 LHS initial design points, so the estimated objective surface remains relatively diffuse. In the middle panel, after several RT evaluations have been added, the acquisition rule begins directing samples toward the upper portion of the feasible region. By the final panel, \(\mu_{13}\) shows a clearly localized high-value region near the upper-right boundary of \(\cX^{\mathrm{DA}}(\cP)\), indicating that BO has identified the most promising part of the feasible space. 

The fact that the maximizer lies on the boundary of the feasible polytope is
expected: the optimal DA allocation often exhausts the available energy
flexibility, so the SoC constraints become binding. Thus, the boundary solution reflects the value of utilizing the battery's
energy capacity, rather than a numerical artifact. The same
binding behavior appears in Figure~\ref{fig:baseline_results} below, where several
final charge and discharge blocks hit the power limits. Additional BO diagnostics, including the corresponding posterior variance surfaces $\sigma_n(\bx)$ and the trace of the regret $\bar{r}_n$ used for early stopping, are reported in Supplementary Figures~\ref{fig:bo_sigma_panels}-\ref{fig:bo_regret_bound}. 

%%%%%%%%%%%%%%%%%%%%%
\subsection{ARBO-DART Algorithm}
\label{sec:adaptive_refinement}
The BO framework for the outer DA problem requires a partition $\cP$. The active blocks of $\cP$ should capture low-price and high-price intervals, while maintaining a balance between flexibility and parsimony: excessively long blocks shift too much dispatch into RT recourse which is costly due to $\gamma$, whereas un-needed blocks diminish the computational advantages of the reduced-dimensional formulation.

To address this tradeoff, we propose a refinement framework that \emph{progressively refines} the partitions. The idea is to begin with a coarse partition, solve the corresponding BO problem, and then refine the partition using a score-based rule that identifies where additional DA commitments are most valuable. At each refinement stage, additional decision variables $x_j$'s are introduced and the BO problem is re-solved, with this process continuing until a stopping criterion is met.

Let \(\cP^{(m)}\) denote the current partition at refinement stage \(m\).
Given \(\cP^{(m)}\), we solve the corresponding fixed-partition BO problem over the feasible set
\(\cX^{\mathrm{DA}}(\cP^{(m)})\) using
Algorithm~\ref{alg:bo_fixed_partition}. This yields the DA
solution and corresponding objective value,
\(\bigl(\bx^{\star}(\cP^{(m)}),\, y^{\star}(\cP^{(m)})\bigr)\),
together with the associated optimal RT policy
\(\widehat{\pi}(\cP^{(m)})
:=
\{\widehat{\pi}_k^{\bx^{\star}(\cP^{(m)})}(\cdot,\cdot)\}_{k=0}^{K-1}\).
This induces an \(\mathbb{F}\)-adapted closed-loop control process, denoted by
\(
\bigl(\widehat{\Delta}_k^{RT,\star}(\cP^{(m)})\bigr)_{k=0}^{K-1},
\)
which is then used to define a scoring rule on the current partition
\(\cP^{(m)}\) so as to identify the time intervals where additional
flexibility is most valuable. Then, we construct a refined partition
\(\cP^{(m+1)}\) by either splitting current blocks or introducing new ones, increasing the number of decision variables. Finally, we re-run Algorithm~\ref{alg:bo_fixed_partition} over the
expanded feasible set \(\cX^{\mathrm{DA}}(\cP^{(m+1)})\). 

\paragraph{Candidate blocks.} To formalize, for a given partition
\(
\cP=\{[a_j,b_j)_{s_j}\}_{j=1}^{d_m}
\)
the active DA blocks occupy the subset
\( 
A(\cP^{(m)}):=\bigcup_{j=1}^{d_m}[a_j,b_j)\subset [0,H).
\label{eq:active_support}
\)
The complement \([0,H)\setminus A(\cP^{(m)})\) corresponds to the set of
non-active intervals, on which the DA decision is zero. Let
\(\mathcal G(\cP^{(m)}):=\{G_1,\ldots,G_r\}\) denote the connected
components of this complement \([0,H)\setminus A(\cP^{(m)})=\bigcup_{\ell=1}^r G_\ell\), where each \(G_\ell\) is a disjoint half-open
interval. In particular, for consecutive active blocks \([a_j,b_j)_{s_j}\) and
\([a_{j+1},b_{j+1})_{s_{j+1}}\), the inactive gap between them is
\(G_\ell=[b_j,a_{j+1})\).  The set of candidate blocks for refinement is
\begin{equation}
\label{eq:candidate_intervals}
\cK(\cP^{(m)})
:=
\{[a_j,b_j):j=1,\ldots,d_m\}\cup \mathcal G(\cP^{(m)}).
\end{equation}
Therefore, \(\cK(\cP^{(m)})\) contains both the active block
intervals and the inactive gap intervals where no decision variable has yet been
assigned. Refining the former results in splitting an existing block \([a_j,b_j)\) into
smaller sub-blocks, while the latter means introducing a new
DA block.

\paragraph{Candidate block scores. } The ranking of the candidate blocks $\mathcal{K}(P^{(m)})$ is guided by the feedback RT recourse
\(
\bigl({\Delta}_k^{RT,\star}(\cP^{(m)})\bigr)_{k=0}^{K-1},
\)
associated with
the current DA solution \(\bigl(\bx^{\star}(\cP^{(m)}),\, y^{\star}(\cP^{(m)})\bigr)\). Intuitively, large RT recourse 
suggests that the current DA parametrization is too coarse. 
We now explain how we assign a score (higher is better) $\cS(\cK)$ to a   candidate
interval \(\cK\in\cK(\cP^{(m)})\). 

\medskip
\noindent
\textit{Case 1: \(\cK=[a_j,b_j)\) is an active block.}
If \(r=l+1\), then the block already has unit length and cannot be split
further. In this case, no admissible cut exists, and we assign it score
zero. If \(r\ge l+2\),  for $\tau = l+1,\ldots, r-1$ let 
\( 
S_{\cK^{}}(\tau):=|L_\tau-R_\tau|,
\label{eq:cut_metric}
\)
with 
\begin{equation}
L_\tau
=
\mathbb E\!\left[\sum_{k=4l}^{4\tau-1}{\Delta}_k^{RT,\star}(\cP^{(m)}) \cdot \Delta t\right],
\qquad
R_\tau
=
\mathbb E\!\left[\sum_{k=4\tau}^{4r-1}{\Delta}_k^{RT,\star}(\cP^{(m)}) \cdot \Delta t\right].
\label{eq:left_right_adjustments}
\end{equation}
where the expectation is calculated via Monte Carlo trajectories. We assign the score
\begin{align}
    \mathcal{S}(\cK) := \max_{\tau\in\{l+1,\ldots,r-1\}} S_{\cK}(\tau),
\end{align}
and if this interval is selected for refinement, the active block $[a_j,b_j)$ is
replaced by the two active sub-blocks $[a_j,\tau^\star(\cK))$ and
$[\tau^\star(\cK),b_j)$ where the locally best cut is
$\tau^\star(\cK)
\in
\argmax_{\tau\in\{l+1,\ldots,r-1\}} S_{\cK(\tau)}$.

\medskip
\noindent
\textit{Case 2: \(\cK=[b_j,a_{j+1})\) is an inactive gap.}
If \(r=l+1\), then no interior cut is available, and we set
\begin{equation}
\mathcal{S}(\cK^{})
:=
\left|
\mathbb E\!\left[\sum_{k=4l}^{4r-1}\widehat{\Delta}_k^{RT,\star}(\cP^{(m)}) \cdot \Delta t\right]
\right|.
\label{eq:gap_score_unit}
\end{equation}
If \(a_{j+1}\ge {b_j}+2\), we apply the same imbalance-based criterion as in
\textit{Case 1}, which yields best cut location
\(\tau^\star(\cK)\) and the two subintervals
\(\bigl[b_{j},\tau^\star(\cK)\bigr)\) and \(\bigl [\tau^\star(\cK),a_{j+1} \bigr)\).
Since \(\cK\) is an inactive gap, only one additional block is
introduced, namely
$\bigl [b_{j},\tau^\star(\cK)\bigr)$ if $
|L_{\tau^\star(\cK)}|
\ge
|R_{\tau^\star(\cK)}|$,
 and \([\tau^\star(\cK),a_{j+1})\) otherwise.

\medskip
For each candidate interval \(\cK\in \cK(\cP^{(m)})\), the procedure above assigns a score \(\cS(\cK)\) together with its associated local refinement. We then sort the candidate intervals based on their scores, labeling as \(\cK^{(m)}_1, \ldots, \cK^{(m)}_{N_m}\), where \(N_m := |\cK(\cP^{(m)})|\), so that
\[
\cS(\cK^{(m)}_1) \ge \cS(\cK^{(m)}_2) \ge \cdots \ge \cS(\cK^{(m)}_{N_m}).
\]
Given a prescribed number \(q_m \in \mathbb N\) of additional DA decision variables to be introduced at refinement stage \(m\), we select the top \(q_m\) candidate intervals $\cK^{(m)}_1, \ldots, \cK^{(m)}_{q_m}$ and add the associated intervals with pre-defined signs $s_j$ (see sign assignment rule below) forming
$[a_j,b_j)$, $j=d_m+1,\ldots,d_{m+1}$, with $d_{m+1}=d_m+q_m$.
The  new partition is
\begin{equation}
\cP^{(m+1)}
=
\cP^{(m)}
\cup
\{[a_j,b_j)_{s_j}\}_{j=d_m+1}^{d_{m+1}}.
\label{eq:partition_update}
\end{equation}

\paragraph{Sign assignment for refined/new block.}
Let 
$
B_k^{\mathrm{RT},\star}(\cP^{(m)})
:=
B_{h(k)}^{\mathrm{DA},*}(\cP^{(m)})
+
\Delta_k^{\mathrm{RT},\star}(\cP^{(m)})
$, denote the $K$-dimensional RT charging strategy associated with $\mathbf{B}^{\mathrm{DA},*}(\cP^{(m)}) \equiv \bx^{\star}(\cP^{(m)})$.
For each  \([a_j,b_j)\), \(j=d_m+1,\ldots,d_{m+1}\), we set $s_j$
according to the sign of the expected total power output over that
interval:
\begin{equation}
s_j= \sign( 
\sum_{k=4a_j}^{4b_j-1}
\mathbb E\!\left[
B_k^{\mathrm{RT},\star}(\cP^{(m)})
\right] 
),
\label{eq:sign_rule}
\end{equation}

\paragraph{Stopping rule.}
The candidate scores $\cS(\cK)$ have units of MWh, since they are based on cumulative RT adjustments. We therefore stop refinement once the largest remaining RT adjustment imbalance is small relative to the battery capacity $I_{\max}$. Given a refinement tolerance $\varepsilon_{\mathrm{refine}}>0$, the adaptive refinement procedure terminates once
\(
\cS(\cK^{(m)}_1) \leq \varepsilon_{\mathrm{refine}}\cdot I_{\max},
\)
where $\cK^{(m)}_1$ is the highest-ranked candidate interval at stage $m$. 

\paragraph{Pruning of inactive blocks.}
As refinement progresses, some active DA blocks in the current partition may be assigned zero amplitude by the BO solution. To control dimension growth, we remove such inactive blocks from the BO search space at the next refinement stage. i.e; let \(x_j^\star(\cP^{(m)})\) denote the BO-optimal amplitude associated with the active block \(([a_j,b_j),s_j)\in\cP^{(m)}\). An active block is marked as inactive if
\(
x_j^\star(\cP^{(m)}) = 0.
\)
When forming \(\cP^{(m+1)}\), inactive blocks are fixed to zero and excluded from the sampling procedure and the corresponding BO routine, yielding a lower-dimensional search space \(\cX^{\mathrm{DA}}(\cP^{(m+1)})\). The corresponding intervals remain eligible for the refinement scoring rule and may later be split or reintroduced as active blocks through the candidate set construction in \eqref{eq:candidate_intervals}.

% \paragraph{Further search after the stopping rule}
% The BO runs performed within each refinement step are intended primarily to
% guide partition refinement, rather than to solve the outer DA problem to high
% accuracy. Accordingly, relatively coarse BO settings are used during the
% stage. Once the refinement stopping rule is triggered at
% step \(m=M\), the resulting partition \(\cP^{(M)}\) is taken as the
% final DA parametrization. We then perform an additional BO search over
% \(\cX^{\mathrm{DA}}(\cP^{(M)})\) using a larger evaluation budget
% \(n_{\mathrm{term}}>n_{\max}\) and a stricter stopping tolerance
% \(\varepsilon_{\mathrm{term}}<\varepsilon_{\mathrm{tol}}\), yielding the final
% DA recommendation
% \(\bigl(x^\star(\cP^{(M)}),\, y^\star(\cP^{(M)})\bigr)\)
% together with the corresponding RT policy
% \(\{\hat{\pi}_k^{\,\star,x^\star(\cP^{(M)})}(\cdot,\cdot)\}_{k=0}^{95}\).

Algorithm~\ref{alg:iterative_bo_cutting} presents the full ARBO-DART procedure. Starting from a coarse user-selected initial
partition, the method iteratively solves a fixed-partition BO problem, analyzes
the resulting RT recourse, and selectively enriches the DA parameterization over
intervals where additional flexibility appears most valuable. This refinement
continues until the stopping criterion is triggered.

\begin{remark}
        At stage \(m\), we warm-start the BO routine by re-using all previously collected input--output pairs from stage \(m-1\), in addition to \(n_0\) newly generated design points. That is, the initial BO dataset at refinement stage \(m\) is formed from \(\cD^{(m-1)}\) together with \(n_0\) new evaluations in \(\cX^{\mathrm{DA}}(\cP^{(m)})\), so that
$
|\cD^{(m)}_{\mathrm{init}}| \leq  |\cD^{(m-1)}| + n_0.
$. The inequality accounts for pruning, since previously evaluated points that are incompatible with the reduced search space are discarded.
This warm-start strategy allows the algorithm to retain information learned at the previous refinement stage, thereby accelerating BO, especially near the \((d_{m-1})\)-dimensional hyperplane naturally embedded in the refined \(d_m\)-dimensional feasible polytope.
\end{remark}
\begin{algorithm}[H]
\caption{ARBO-DART algorithm}
\label{alg:iterative_bo_cutting}
\begin{algorithmic}[1]
\Require Initial partition \(\cP^{(1)}\), \(q_m\) additional variables per refinement, refinement threshold \(\varepsilon_{\mathrm{refine}}\), BO parameter schedules \((n^{(m)}_0, n^{(m)}_{\max}, \varepsilon^{(m)}_{\mathrm{tol}})_{m=1,\ldots}\)
\For{$m=1,\ldots,$}
    \State Perform BO over \(\mathcal{X}^{\mathrm{DA}}(\cP^{(m)})\) using Algorithm~\ref{alg:bo_fixed_partition} with parameters \((n^{(m)}_0, n^{(m)}_{\max}, \varepsilon^{(m)}_{\mathrm{tol}})\)
    \State Form the candidate set \(\cK(\cP^{(m)})\) as in \eqref{eq:candidate_intervals}
    \State Compute and rank in decreasing order the block-wise scores \(\mathcal{S}(\cK_j^{(m)}), j=1,\ldots,N_m\) via \eqref{eq:gap_score_unit}
    \If{  \( \mathcal{S}(\mathcal{K}^{(m)}_1) < \varepsilon_{\mathrm{refine}} \cdot I_{\max}\)  } $\qquad$ // Stop
        \State Set \(\mathcal{M} \gets m\) and do extra BO iterations up to budget $n_{\max}^{(\cM)}$
        \State \Return \(\cP^{(\mathcal{M})}\), \(\bx^\star(\cP^{(\mathcal{M})})\), \(y^\star(\cP^{(\mathcal{M})})\) and \(\widehat{\pi}(\cP^{(\mathcal{M})})\)
    \EndIf
    \State Select the top \(q_m\) intervals for refinement: $\cK_1^{(m)}, \ldots, \cK_{q_m}^{(m)}$
    \State Assign sign(s) to the refined/new block(s) using \eqref{eq:sign_rule}
    \State Update \(\cP^{(m)}\) to obtain \(\cP^{(m+1)}\) via \eqref{eq:partition_update}; prune inactive blocks
\EndFor
\end{algorithmic}
\end{algorithm}

%%%%%%%%%%%%%%%%%%%%%
\section{Numerical Results}
\label{sec:numerical_results}

\subsection{ Case Study 1: Zero-Mean DART Spread}
In this section, we apply ARBO-DART
to a realistic problem instance. We use the DA price vector and mean-reverting RT
price process \eqref{eq:rt_factor_price} from Section~\ref{sec:problem_setup},
and consider a 4-hour battery with 95\% efficiency. The continuation-value and policy emulators for the
RT recourse layer are trained with $\mathcal{N}_{\mathrm{loc}}=\mathcal{N}_{b} = 100$ LHS design points per step.
To initialize ARBO-DART, we use a simple domain-informed heuristic, starting off with a 2-block
partition with one 4-hour charging block around the lowest-price period and one
4-hour discharging block around the highest-price period, analogous to TB4. At each refinement
stage $m$, we add $q_m=3$ decision variables and stop refinement when the
cumulative RT adjustment falls below $\varepsilon_{\mathrm{refine}}\cdot I_{\max}=0.40$, $10\%$ of the maximum
energy capacity. The BO subroutine is initialized with
\( 
n_0^{(m)}=\left\lfloor 6\sqrt{d_m}\right\rfloor
\)
LHS design points. For all stages preceding the final stage \(\mathcal{M}\) where the stopping criterion is triggered, we use a fixed BO regret tolerance
\( 
\epsilon_{\mathrm{tol}}^{(m)}=0.1
\) 
and a fixed evaluation budget
\(
n_{\max}^{(m)}=20 + n_0^{(m)}.
\)
At the final stage \(\mathcal{M}\), we instead use the adaptive budget
\( 
n_{\max}^{(\mathcal{M})}
=
n_0^{(\mathcal{M})}
+
5d_{\mathcal{M}},
\)
which allows additional exploration of the final partition before selecting the reported candidate as the final DA profile.  In
Section~\ref{sec:ablation}, we examine the sensitivity of ARBO-DART to the initial block length, seed, number of cuts and BO budgeting. 
The full parameter set is reported in Table~\ref{tab:baseline_params}. With the given SHADOw-GP settings, each RT evaluation takes about 1 minute on a laptop. 

\begin{table}[!htb]
\centering
\caption{Parameters for Case Study 1.}
\label{tab:baseline_params}
\begin{tabular}{llll}
\hline
Category & Parameter & Value & Description \\
\hline

\multirow{3}{*}{\shortstack[l]{RT Process\\Dynamics}}
& $\kappa$ & $0.2$ & Mean reversion rate \\
& $\sigma$ & $1.0$ & Volatility \\
& $\lambda$ & $0.05$ & OU factor scale \\
\hline

\multirow{4}{*}{\shortstack[l]{BESS\\Characteristics}}
& $\eta$ & $0.95$ & Charging/discharging efficiency \\
& $\bar B$ & $1$ MW & Charging/discharging power limit \\
& $I_{\max}$ & $4$ MWh & Energy capacity \\
& $I_0$ & $1.0$ MWh & Initial and target SoC \\
\hline

\multirow{2}{*}{RT Objective}
& $\gamma$ & $0.2$ & RT cost \\
& $\rho$ & $100$ & Terminal SoC penalty \\
\hline

\multirow{2}{*}{\shortstack[l]{RT Solver\\Training}}
& $\mathcal{N}_{\mathrm{loc}}=\mathcal{N}_{\mathrm{b}}$
& $100$
& LHS design points for $\widehat{Q}_k$ and $\widehat{\pi}^{*}_k$ \\
& $\mathcal{N}_{\mathrm{rep}}$
& $10$
& Replications per LHS design point for $\widehat{Q}_k$ \\
\hline

\multirow{6}{*}{\shortstack[l]{ARBO-DART\\Stages}}
& $\mathcal{P}^{(1)}$
& $\{[9,13)_{+},[16,20)_{-}\}$
& Initial partition, $d_1=2$ \\
& $q_m$
& $3$
& Added variables per refinement stage \\
& $\varepsilon_{\mathrm{refine}}$
& $0.10$
& Refinement stopping threshold \\
& $n_0^{(m)}$
& $\lfloor 6\sqrt{d_m}\rfloor$
& Initial LHS design size \\
& $\epsilon_{\mathrm{tol}}^{(m)}$
& $0.1\sqrt{2/d_m}$
& BO stopping tolerance \\
& $n_{\max}^{(m)}$
& $n_0^{(m)}+20$
& Maximum RT  evaluations per stage \\
\hline
\end{tabular}
\end{table}

\begin{figure}[!htb]
    \centering
    \hfill
    \begin{subfigure}[t]{0.64\textwidth}
        \centering
        \includegraphics[width=\linewidth]{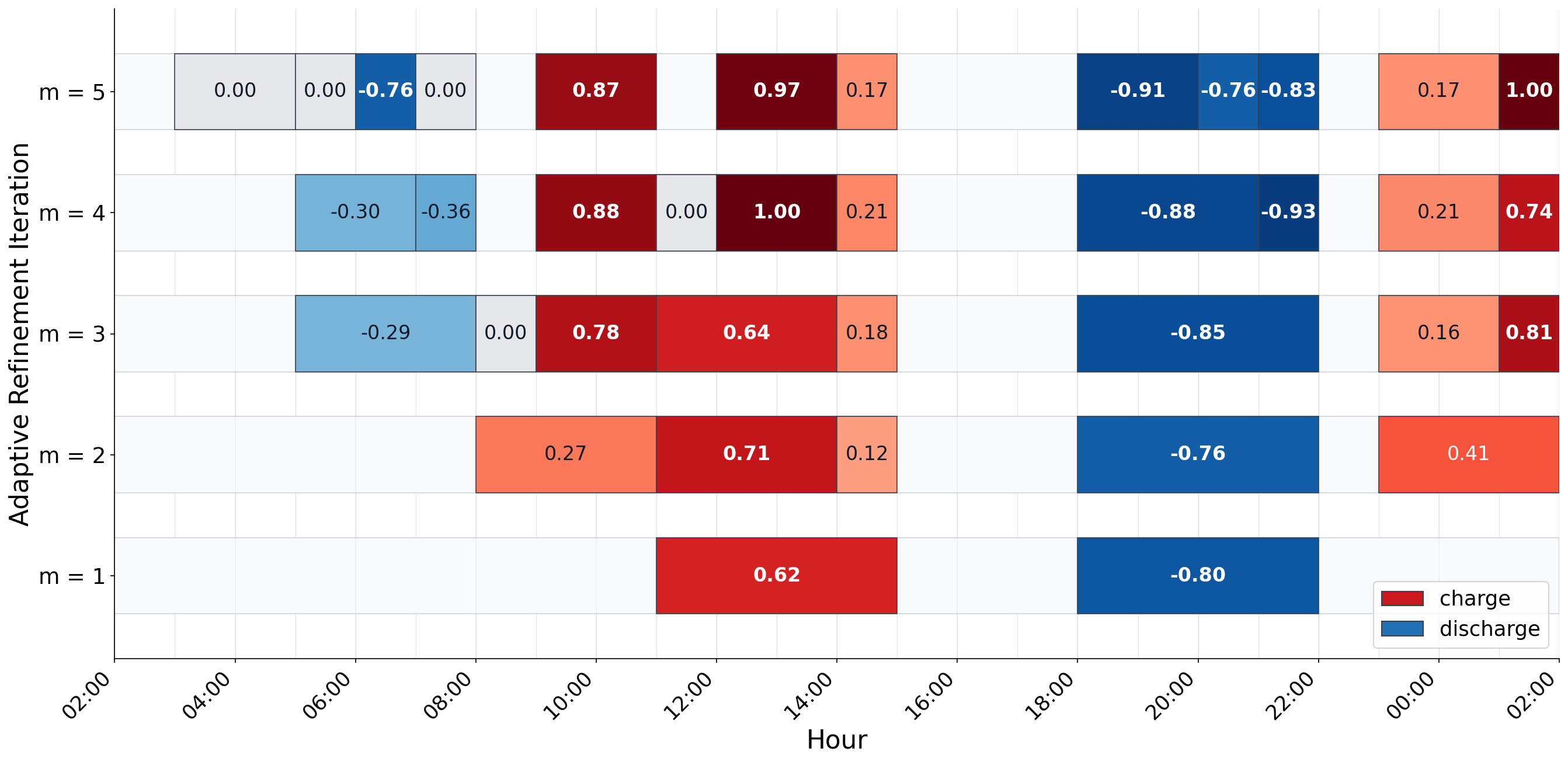}
        \caption{Adaptive DA partition evolution.}
        \label{fig:baseline_partition_evolution}
    \end{subfigure}
    \vspace{0.25cm}
    \begin{subfigure}[t]{0.34\textwidth}
        \centering
        \includegraphics[width=\linewidth]{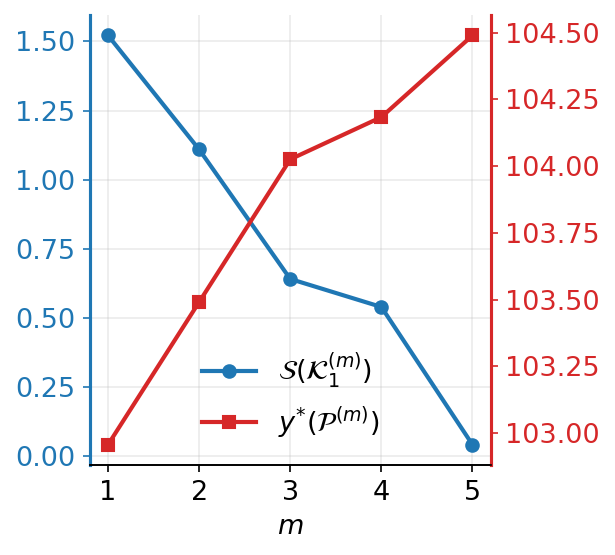}
        \caption{Refinement score and objective value.}
        \label{fig:baseline_score_profit}
    \end{subfigure}

    \begin{subfigure}[t]{0.5\textwidth}
        \centering
        \includegraphics[width=\linewidth]{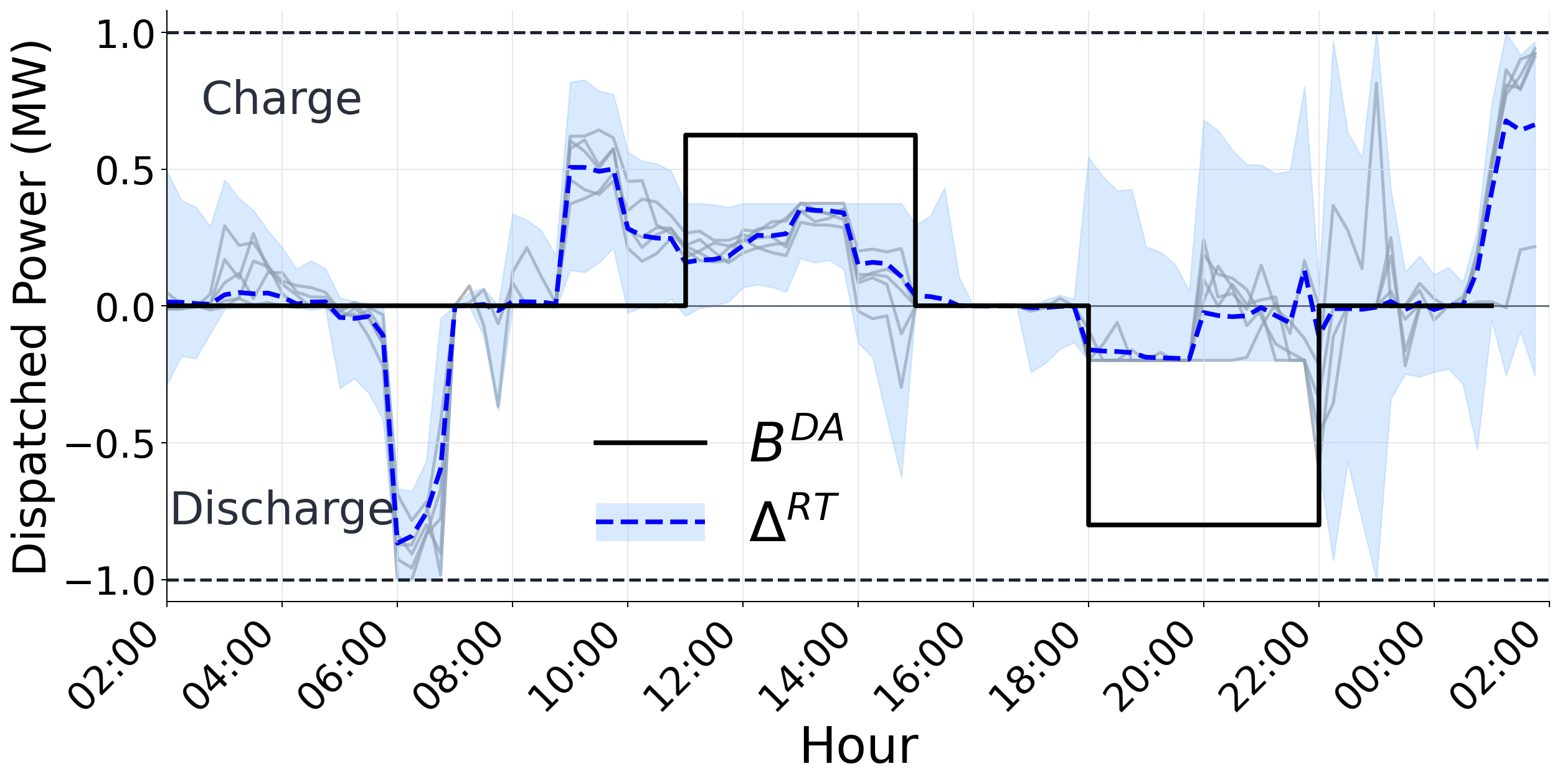}
        \caption{DA profile with RT adjustment at $m=1$}
        \label{fig:baseline_partition_p1}
    \end{subfigure}
    \begin{subfigure}[t]{0.48\textwidth}
        \centering
        \includegraphics[width=\linewidth]{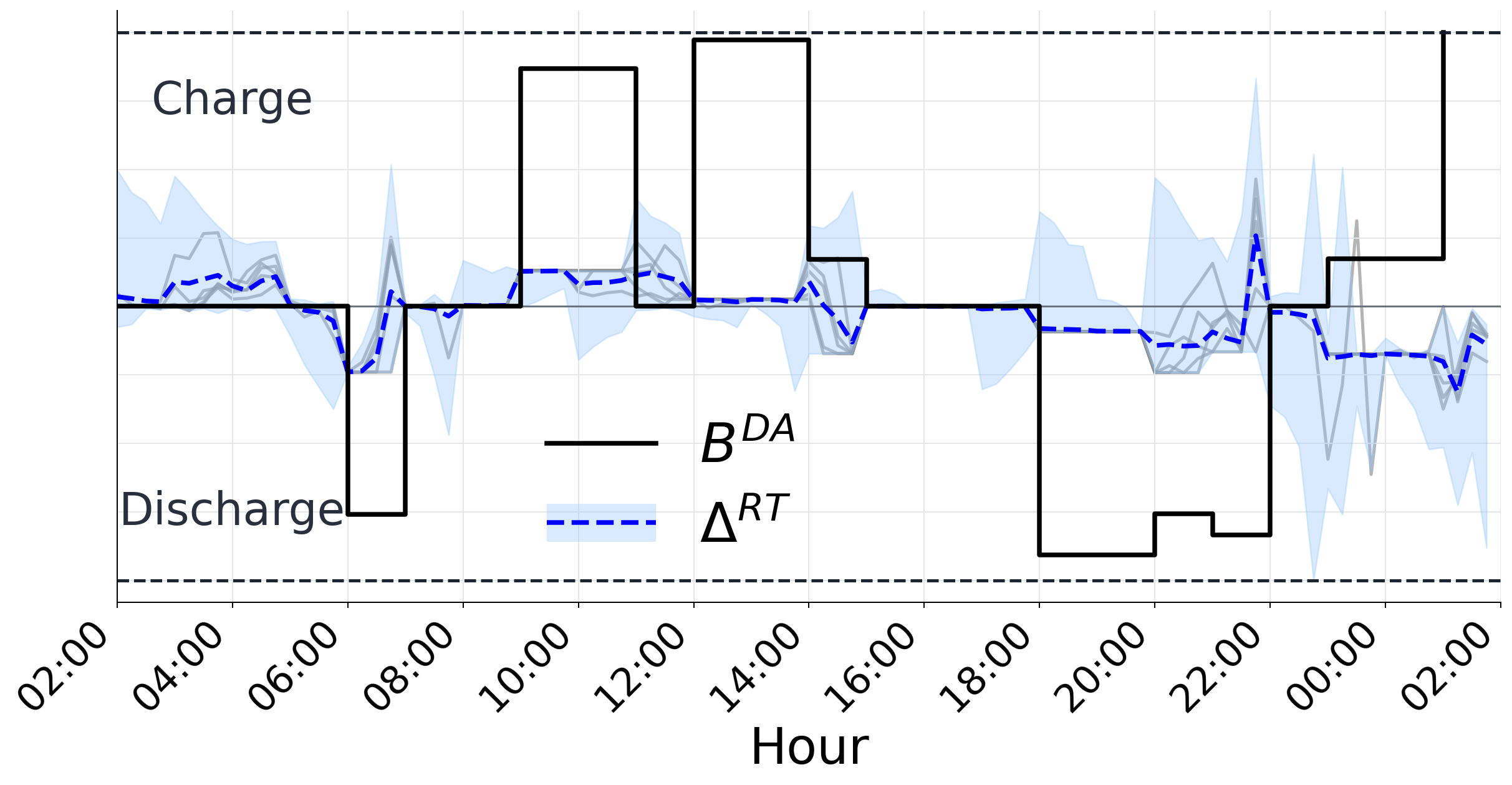}
        \caption{DA profile with RT adjustment at $m=5$}
        \label{fig:baseline_partition_p5}
    \end{subfigure}

\caption{
\emph{Top left}: refinement of the DA partition \(\cP^{(m)}\) across stages $m$,
with block location indicating time, and color
indicating sign and magnitude of DA dispatch.
\emph{Top right}: leading refinement score
\(\cS(\cK^{(m)}_1)\) and optimized objective value
\(y^\star(\cP^{(m)})\) across refinement stages $m=1,\ldots,5$.
\emph{Bottom}: optimized DA profiles and expected RT adjustments under the initial
partition \(\cP^{(1)}\) (\emph{left}) and final partition \(\cP^{(5)}\) (\emph{right}).
Black step functions show \(\mathbf{B}^{\mathrm{DA},\star}(\cP^{(m)})\). Dashed blue curves
show \(\mathbb E[\Delta^{\mathrm{RT},\star}(\cP^{(m)})]\) with the bands indicating the corresponding 99-th percentile range. We moreover show 5 sample trajectories of RT recourse $\Delta^{\mathrm{RT},\star}(\cP^{(m)})$. }

    \label{fig:baseline_results}
\end{figure}

The left panel of Figure~\ref{fig:baseline_results}(c) illustrates the optimized DA profile and corresponding RT adjustments after applying BO Algorithm~\ref{alg:bo_fixed_partition} to $\cP^{(1)}$. Accounting for the \(95\%\) charging and discharging efficiency, the DA strategy charges the battery from its initial SoC of \(1\) MWh up to \(\simeq 3.36\) MWh during the \(11{:}00\)--\(15{:}00\) window, and then discharges during \(18{:}00\)--\(22{:}00\), reducing the SoC toward \(0\) MWh. During RT recourse, the BESS charges between \(00{:}30\) and \(02{:}00\) to return its terminal SoC toward the \( I_0 = 1\) MWh target.
 Because \(\cP^{(1)}\) is coarse,  much of the finer intraday variation is
left to the RT controller. The sizable RT adjustments, including during periods
with zero DA allocation, show that (as expected) the initial partition is inadequate.
Based on these adjustments, the scoring rule then introduces two charging blocks, 8:00--11:00 and 23:00--02:00 and splits the mid-day charging block 11:00--16:00 into 11:00--14:00 and 14:00--15:00. This produces the next partition 
\(
\cP^{(2)}
=
\{[6,9)_{+},\; [9,12)_{+},[12,13)_{+}, [16,20)_{-},[21,24)_{-}\}.
\)

Figure~\ref{fig:baseline_results}(a) shows the sequence of DA partitions
introduced by ARBO-DART. After each refinement, the DA profile is re-optimized over the
updated partition, allowing the newly introduced blocks to absorb the
adjustment that was previously handled by RT recourse. This process continues
until the stopping rule is triggered at \(\mathcal{M}=5\), as indicated by the declining
refinement score $\cS(\cK^{(m)}_1)$ in Figure~\ref{fig:baseline_results}(b).
Furthermore,
Figure~\ref{fig:baseline_results}(a) shows the pruning step in action at
\(m=3,4\), where blocks with zero DA amplitude
are dropped from the next refinement stage.  Figure~\ref{fig:baseline_results}(d) shows the optimized DA profile and expected RT adjustment after ARBO-DART is applied to the final partition \(\cP^{(5)}\) which yields a DA profile with nine non-zero blocks. We observe that the resulting DA profile $\mathbf{B}^{\mathrm{DA}, (5)}$ slightly over-charges during 22:00--01:00 in order to return the battery toward the initial SoC by the end of the horizon; this excess DA position is then corrected by RT recourse. 
This is further evident from the expected total energy bought and sold, defined as
\[
E_{\pm}
:= \pm 1 \cdot 
\mathbb{E}\!\left[
\sum_{k=0}^{K-1} \eta^{\pm1}
B_k^{\mathrm{RT}}
\mathbb{1}_{\{B_k^{\mathrm{RT}}\gtrless 0\}}
\Delta t
\right].
\] By symmetry, due to the identical initial and terminal SoC conditions, \(E_+=E_{-}=4.98\,\mathrm{MWh}\) where DA trading accounts for \(99.36\%\) of the energy bought and \(87.76\%\) of the energy sold. Additional policy diagnostics of SoC, alongside the dispatched power for \(\cP^{(1)}\), \(\cP^{(3)}\),
and \(\cP^{(5)}\) are provided in Supplementary Figure~\ref{fig:supp_three_policies_viz}. 
\paragraph{RT trading friction \(\gamma\):} The cost $\gamma$ of RT adjustments 
dictates the overall split between the DA and RT layers. Higher $\gamma$ makes DA actions preferable, while $\gamma \simeq 0$ gives no benefit to dispatching DA and hence shifts all attention to RT. To examine the sensitivity of the ARBO-DART solution to the RT recourse penalty, we vary \(\gamma\in\{0.1,0.2,0.4\}\). In particular, we study how \(\gamma\) changes the magnitude of RT recourse, the resulting DA profile, and the terminal BO dimension. Because RT adjustments are generally close to zero, average RT recourse is not a good metric.  We therefore report the average \(L^2\) norm:
\begin{align}\label{eq:l2-norm}
\|\Delta^{\mathrm{RT}}\|_2
:=
\mathbb{E}\left[
\sqrt{\sum_{k=0}^{K-1}
\left(\Delta_k^{\mathrm{RT}}\right)^2}
\right],
\end{align}
measured in MW. This metric summarizes the expected amplitude of RT recourse over the day, with a larger $L^2$-norm indicating greater reliance on RT adjustments.

As \(\gamma\) increases, large RT adjustments become more expensive, leading to the decrease in \(\|\Delta^{\mathrm{RT}}\|^2\) reported across the corresponding rows of Table~\ref{tab:curve_pnl_decomposition}. The reduced flexibility of RT recourse also changes the optimized DA profile, which becomes organized into fewer, more contiguous charging and discharging blocks. Consequently, the terminal BO dimension \(d_{\mathcal{M}}\) decreases monotonically with \(\gamma\), as finer DA partitioning provides less value when RT corrections are more penalized; see Supplementary Figure~\ref{fig:supp_gamma_variation}.

\begin{table}[!htb]
\centering
\caption{Economic and energy decomposition of ARBO-DART across case studies. PnL values are reported in dollars; aggregate dispatch percentage is from \eqref{eq:cycle} and the $L^2$-norm of $\Delta^{\mathrm{RT}}$ is from \eqref{eq:l2-norm}.}
\label{tab:curve_pnl_decomposition}
\begin{tabular}{ccc|rrrrrr}
\hline
\textbf{CS} &
\(\boldsymbol{I_{\max}}\) &
\(\boldsymbol{\gamma}\) &
\shortstack{\textbf{DART} \\ \textbf{PnL}}  &
\shortstack{\textbf{DA-only} \\ \textbf{PnL}} &
\shortstack{\textbf{DA} \\ \textbf{TB}} &
\shortstack{\textbf{ADP} \\ (\%)}&
\shortstack{ \textbf{RT Norm} \\ \(\boldsymbol{\|\Delta^{\mathrm{RT}}\|_2}\)} &
\shortstack{\textbf{Final} \\  $\boldsymbol{d_\mathcal{M}}$} \\
\hline\hline
1 & 4 & 0.1 & 104.53  & 101.78 & 76.77 & 127 & 2.38 & 14 \\
1 & 4 & 0.2 & 104.50 & 101.78 & 76.77 & 125 & 1.29 & 12\\
1 & 4 & 0.4 & 104.49 & 101.78 & 76.77 & 125 & 0.69 & 10 \\ \hline 
2 & 4 & 0.2 & 113.57 & 101.78 & 76.77 & 133 & 3.46 & 12 \\
3 & 4 & 0.2 & 72.65  & 66.59  & 38.83 & 180 & 3.10 & 11 \\
4 & 4 & 0.2 & 140.53 & 138.38 & 132.15 & 102 & 1.41 & 8 \\ 
4 & 2 & 0.2 & 79.82   & 78.43  & 75.20 & 105 & 1.06 & 8\\
\hline
\end{tabular}
\end{table}
We further examine how \(\gamma\) affects the expected  DART PnL, \(\mathbb{E} [\Pi_{K-1}^{\mathrm{DART}}]\), and the aggregate battery charge/discharge over the operating day \(D\), akin to a daily battery cycling metric. We define
the aggregate dispatch percentage (ADP)
\begin{align}\label{eq:cycle}
\mathrm{ADP}
:=
100 \times \frac{E_{\pm}}{I_{\max}},
\end{align}
as the total charged or discharged energy during the operating day expressed as a percentage of the battery energy capacity. As shown in Table~\ref{tab:curve_pnl_decomposition}, the expected DART PnL changes marginally across the considered values of $\gamma$. The ADP also remains relatively stable at approximately $125\%-127\%$, interpreted as equivalent to 1.25 full charge-discharge cycles. 

%%%%%%%%%%%%%%%%%%%%%
\subsection{Case Study 2: Systematic DART Spread}\label{sec:dart-spread}

\begin{figure}[!htb]
    \centering
    \begin{subfigure}[t]{\textwidth}
        \centering
        \includegraphics[width=0.60\linewidth]{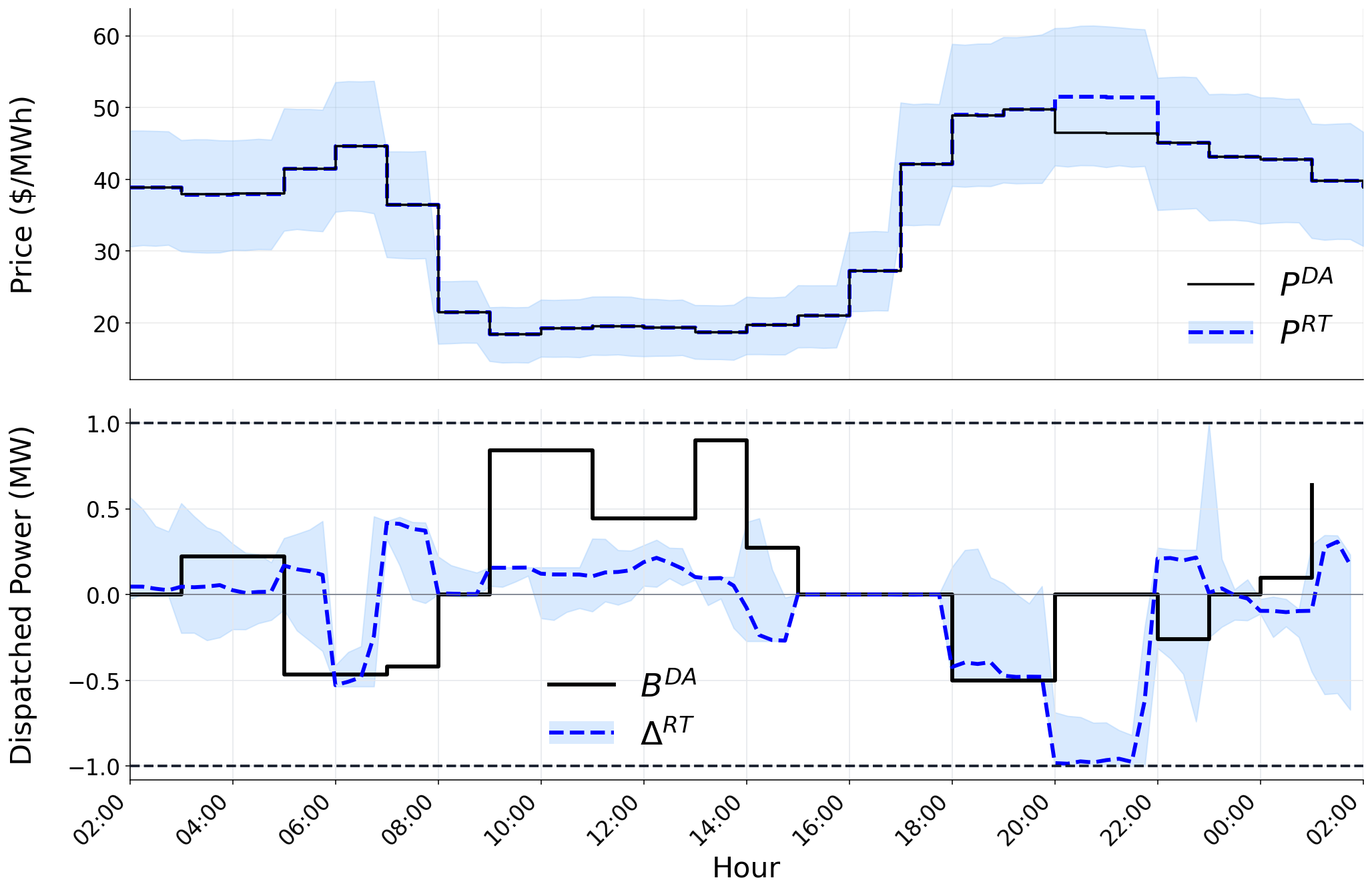}
        \label{fig:biased_da_policy_panel}
    \end{subfigure}

    \vspace{0.15cm}

    \begin{subfigure}[t]{\textwidth}
        \centering
        \includegraphics[width=0.60\linewidth]{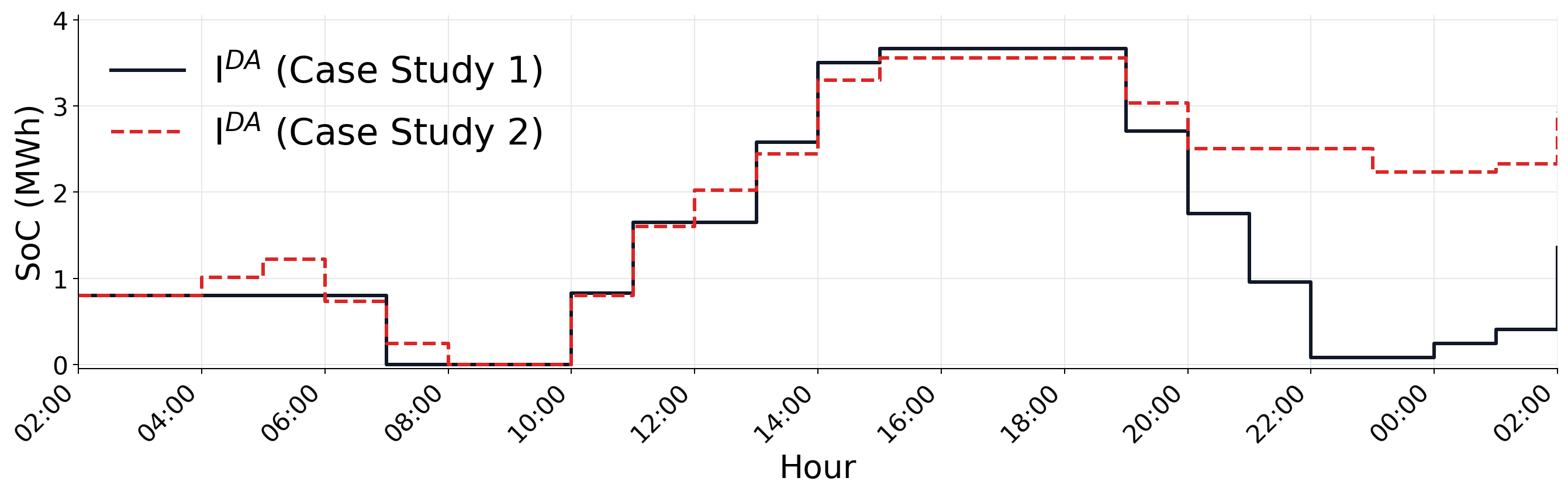}
        \label{fig:biased_inventory_comparison_panel}
    \end{subfigure}

    \caption{Case Study 2: Systematic DART spread. \emph{Top panel:} RT and DA price profiles; $\mathbf{P}^{\mathrm{RT}}$ follows \eqref{eq:biased_rt_factor_price} with an upward bias during 2 evening hours. \emph{Middle
    panel:} resulting optimized DA bids \(\mathbf{B}^{\mathrm{DA},\star}(\cP^{(5)})\) (black step function) and RT adjustment 
    \(\mathbb{E}[{\Delta}^{\mathrm{RT},\star}(\cP^{(5)})]\) (blue) with its $99$-th percentile band. \emph{Bottom:} DA SoC $\mathbf{I}^{\mathrm{DA}}$ under Case Studies 1 and 2, highlighting the impact of a non-zero DART spread on DA SoC headroom.}
    \label{fig:biased_case_results}
\end{figure}

The baseline RT price model \eqref{eq:rt_factor_price} assumes that deviations from the DA price are
mean-zero, so that the DART spread is zero in expectation. In this case study,
we consider a setting with a systematic RT price premium during a selected
delivery window. This represents a predictable DART spread
that may arise from recurring congestion, forecast errors, or scarcity
conditions. To this end, we preserve the mean-zero OU
factor model for stochastic RT fluctuations, but add  a deterministic bias during 20:00--22:00. Over this window, which is right after the DA price peak, the expected RT price is shifted upward by \(\$5/\mathrm{MWh}\) in each delivery hour, producing a cumulative arbitrageable DA--RT spread of \$10:
\begin{equation}
P_k^{\mathrm{RT}}
=
P_{h(k)}^{\mathrm{DA}}
+
b_{h(k)}
+
\lambda P_{h(k)}^{\mathrm{DA}}Y_k,
\qquad k=0,\ldots,K-1, \qquad b_h =
\begin{cases}
5, & h \in \{18,19\},\\
0, & \text{otherwise},
\end{cases}
\label{eq:biased_rt_factor_price}
\end{equation}
where \(Y_k\) follows the same mean-zero OU factor dynamics as 
\eqref{eq:ou_factor_dynamics}. 
The top panel of Figure~\ref{fig:biased_case_results} illustrates the $99$-th percentile band and mean of RT simulations generated by
\eqref{eq:biased_rt_factor_price}, alongside the DA price.

We then apply the ARBO-DART algorithm using the same configuration as in the baseline case. The algorithm again terminates at \(\mathcal{M}=5\). The middle panel of Figure~\ref{fig:biased_case_results} shows the optimized DA profile and the expected RT adjustment, together with the \(99\%\) percentile band, after BO is applied to the final partition \(\mathcal{P}^{(5)}\). During the biased 20:00--22:00 window, the optimized DA commitment is zero, while the expected RT adjustment is close to full discharging. By contrast, the Case 1 policy schedules discharge through the DA market over the same interval, as shown in the bottom-right panel of Figure~\ref{fig:baseline_results}. In this case study, \(E_{\mathrm{bought}}=E_{\mathrm{sold}}=5.57\,\mathrm{MWh}\). DA trading accounts for \(88.23\%\) of the energy purchased but only \(51.56\%\) of the energy sold. The latter share is substantially lower than in Case 1 because the RT price bias creates an incentive to defer discharge to the RT market. This shift is also visible in the bottom panel of Figure~\ref{fig:biased_case_results}, which compares the corresponding DA SoC trajectories. The two trajectories remain similar throughout the morning and afternoon but begin to diverge around 20:00, when \(\mathbb{E}[P^{\mathrm{RT}}]\neq P^{\mathrm{DA}}\). Furthermore, compared with Case Study 1, this case exhibits larger RT recourse, as expected; see Table~\ref{tab:curve_pnl_decomposition}. Profit increases by \(\$9.07\); compared to  the \( \$10\) arbitrage expected DART spread, the \(93\) cents loss is attributable to the RT recourse penalty \(\gamma\).

%%%%%%%%%%%%%%%
\subsection{Alternative Price Curves}
\label{ssec:alt_curves}
To validate the stability of our algorithm, we next evaluate it on two alternative DA
price curves.

\begin{figure}[!htb]
    \centering
    \begin{tabular}{cc}
    \includegraphics[
        width=0.48\linewidth,
        trim=0.2in 0.15in 0.2in 0.2in        
    ]{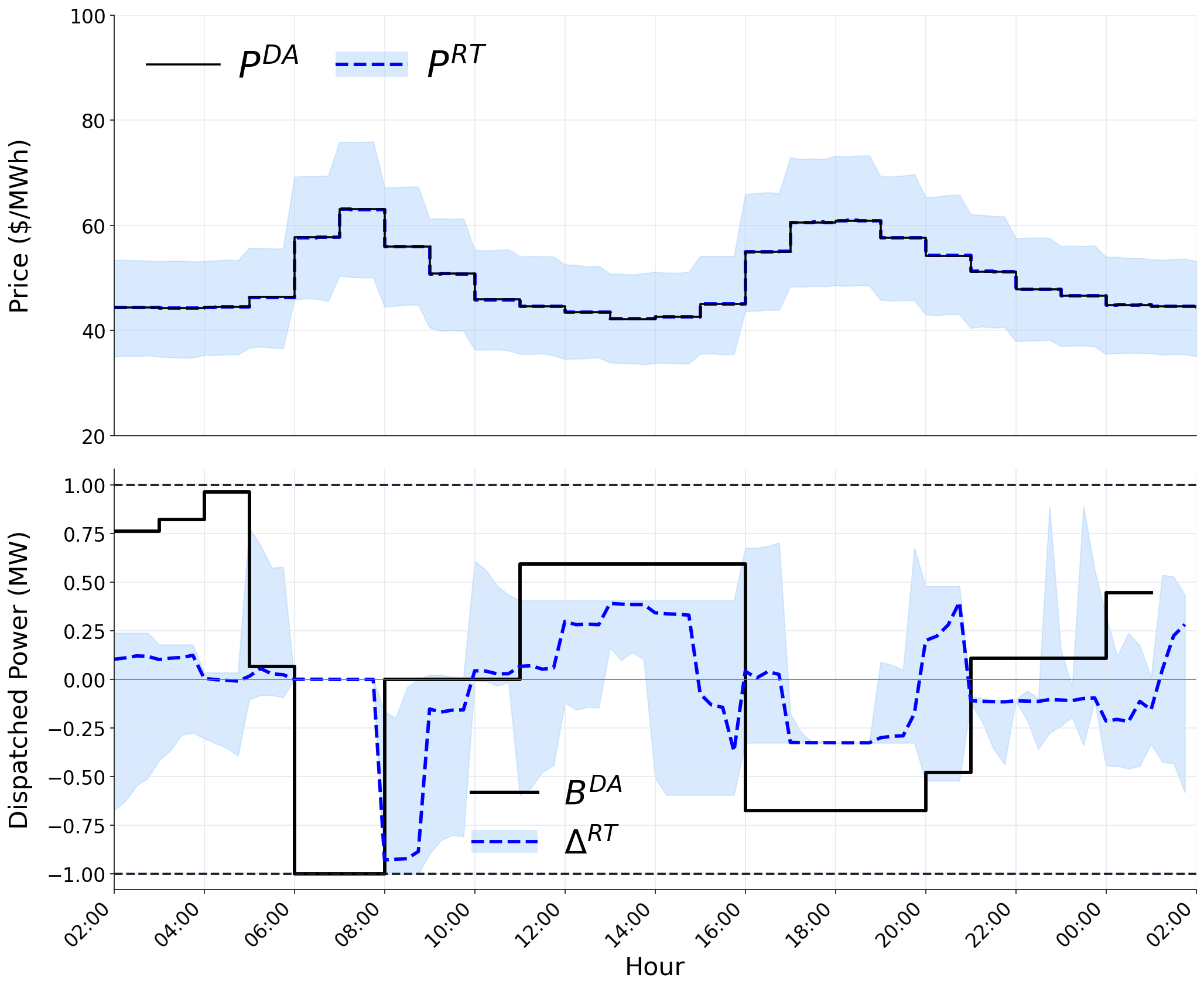}
    &
    \includegraphics[
        width=0.47\linewidth,
        trim=0.5in 0.15in 0.02in 0.02in
    ]{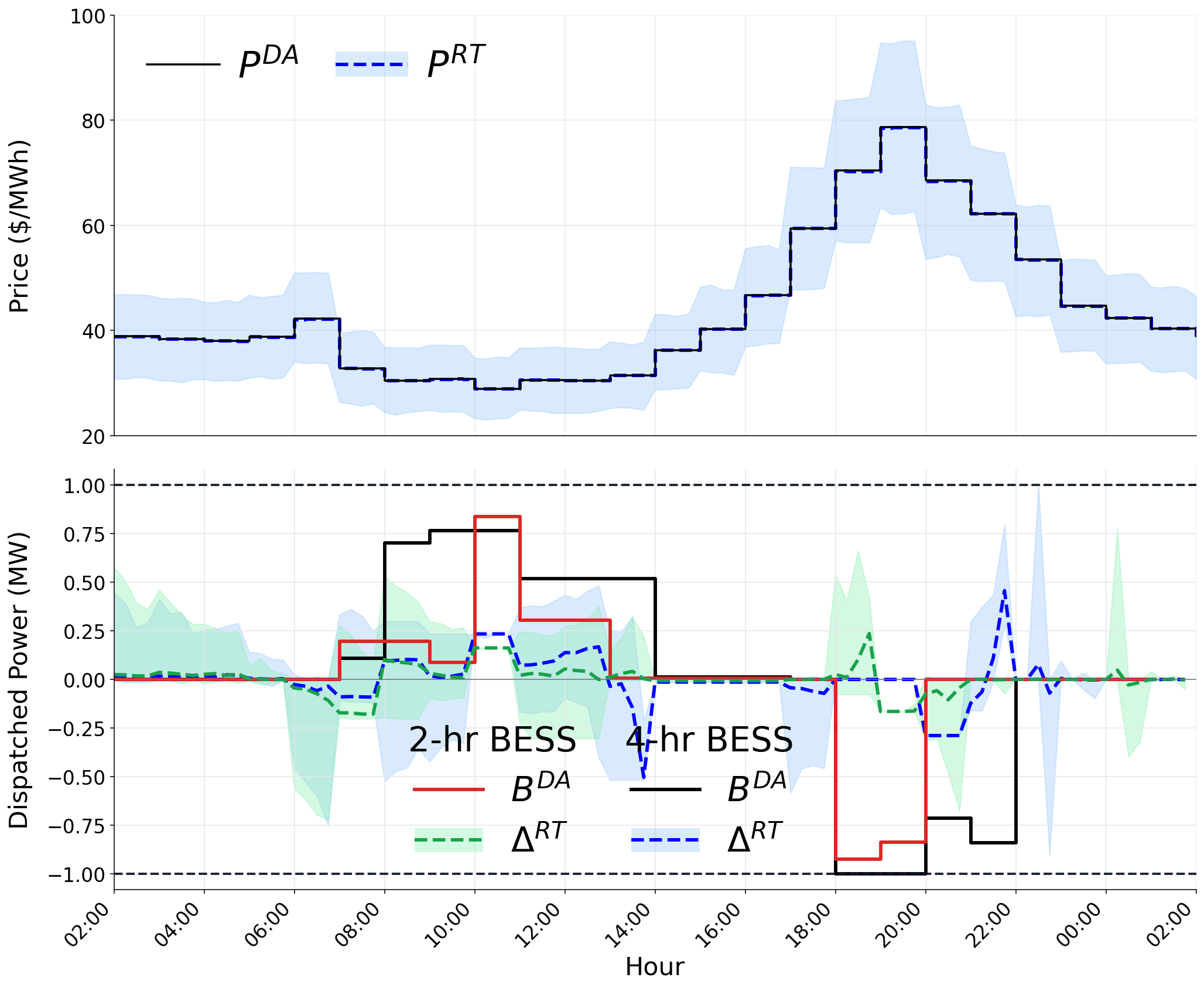}
    \\
    Case Study 3 & Case Study 4
    \end{tabular}
    \caption{Case Studies 3 (\emph{left}) and 4 (\emph{right}): optimized DA profile and expected RT adjustment.
    In the bottom panels, the black step functions show
    \(\mathbf{B}^{\mathrm{DA},\star}(\cP^{(\mathcal{M})})\), while the blue curves show
    \(\mathbb{E}[\Delta^{\mathrm{RT},\star}_k(\cP^{(\mathcal{M})})]\)
    with their 99th-percentile bands. The dashed horizontal lines indicate the maximum
    charge and discharge limits, \(\pm\bar{B}\). In the bottom-left panel, we compare
    the optimized DA profiles \(\mathbf{B}^{\mathrm{DA},\star}\) and RT adjustments
    \(\Delta^{\mathrm{RT},\star}\) for 4-hour (\(I_{\max}=4\) MWh) and 2-hour
    (\(I_{\max}=2\) MWh) BESS.}
    \label{fig:CS_3_4_results}
\end{figure}

\textbf{Case Study 3:} Left panel of Figure~\ref{fig:CS_3_4_results} shows a flatter DA curve with  mild intraday variation. There are two energy arbitrage opportunities: one in the morning and one
in the evening, leading to two charge-discharge cycles. 
The experimental setup is the same as in
Table~\ref{tab:baseline_params} and the initial DA partition is chosen as
\(
\cP^{(1)}=\{[0,4)_{+},[4,8)_{-}\},
\)
which initializes the search around the morning price valley and peak. The refinement terminates after \(\mathcal{M}=4\) 
stages with score $\mathcal{S}({\mathcal{K}^{(\mathcal{M})}_1}) = 0.375$. The bottom left panel of Figure~\ref{fig:CS_3_4_results} illustrates the final ARBO-DART policy. During the first cycle, from 02:00 to 09:00, most charging and discharging activity is allocated directly to the DA market. By contrast, during the second cycle, from 11:00 to 21:00, the policy uses a more balanced allocation between the DA market and RT recourse.

The presence of two cycles results in a higher ADP and $L^2$ norm of RT recourse, as reported in Table~\ref{tab:curve_pnl_decomposition}. The milder price variation results in a profit of \(\$72.65\). For comparison, the DA-only optimizer achieves \(\$66.59\), whereas the DA TB4 baseline achieves only \(\$38.83\). The substantially lower performance of DA TB4 arises because its fixed four-block structure captures only one of the two trading cycles. 

\textbf{Case Study 4:} Right panel of Figure \ref{fig:CS_3_4_results} shows a DA curve with with a relatively flat
overnight and midday profile, followed by a steep evening price peak. This profile is dominated by a single large evening opportunity around 19:00, with TB2 of around \$70; only a single charge-discharge cycle is expected.
We initialize with
\(
\cP^{(1)}=\{[7,11)_{+},[16,20)_{-}\}
\)
and to examine the role of the
terminal condition, we set the initial and target terminal SoC to zero,
\(I_0=0\). ARBO-DART terminates after \(\mathcal{M}=3\) refinement stages, earlier
than in the other case studies, with cutting score $\mathcal{S}(\mathcal{K}^{(\mathcal{M})}_1) = 0.122$. We attribute this earlier termination to the simpler one-cycle structure and the zero SoC terminal condition. As shown in the 
bottom right panel of Figure~\ref{fig:CS_3_4_results}, the DA strategy has a
simple charge--discharge structure: it charges over 
07:00--14:00, and discharges over 
18:00--22:00 during the evening price peak. Unlike the previous case studies, we do not observe end-of-day charging. In this case study, the final BO dimension is \(d_{\mathcal{M}}=8\); see Table~\ref{tab:curve_pnl_decomposition}. This dimension is lower than in the previous case studies because of the simpler terminal condition and economic structure of the price curve. 
As expected, the estimated number of battery cycles is approximately \(1\). The reported PnL is \(\$140.53\), which is \(\$2.15\) higher than that of the DA-only optimizer. 

\paragraph{Battery duration.}
Battery duration is a key characteristic of BESS; longer duration batteries are more suitable for flatter charge-discharge cycles. We repeat Case Study 4  with a 2-hour battery by setting \(I_{\max}=2\) and initializing with two 2-hour blocks,
\(
\cP^{(1)}=\{[7,9)_{+},[16,18)_{-}\}
\). ARBO-DART terminates after \(\mathcal{M}=3\) refinement
stages with cutting-score $\mathcal{S}(\mathcal{K}^{(\mathcal{M})}_1) = 0.079$. As shown in the bottom right panel of
Figure~\ref{fig:CS_3_4_results}, smaller battery capacity leads to a shorter charge--discharge structure. In particular, the evening DA discharge
period shrinks from 18:00--22:00 in the 4-hour case to
18:00--20:00 in the 2-hour case. As a result, the 2-hour PnL drops from \(\$140.53\) to \(\$79.82\) in Table \ref{tab:curve_pnl_decomposition}, a reduction of approximately \(43\%\), reflecting the 50\%-smaller energy capacity available to exploit the evening price peak.

%%%%%%%%%%%%%%%%%%%%%%%%%
\subsection{Comparators}\label{ssec:comparator}

To have external validation of ARBO-DART, we implement three comparators.  The comparison is conducted for Case studies 1 and 4, under the default setting \(I_{\max}=4\) and \(\gamma=0.2\). In CS1, the initial and terminal SoC are both set to \(1\,\mathrm{MWh}\), whereas in CS4 they are both set to \(0\,\mathrm{MWh}\). 

The first comparator is a 24-dimensional BO solver that directly optimizes the full hourly DA dispatch profile in \eqref{eq:dart_exact}. This represents the natural brute-force approach for optimizing the black-box DART objective, dropping adaptive refinement. Next, we implement a comparator that solves a fixed 12-dimensional setup whereby the DA dispatch is constrained to remain constant over each consecutive two-hour interval. The resulting lower-dim.~search space is easier for BO to explore. These two comparators assess the benefit of adaptive refinement; comparing between the 12- and 24-dim.~formulations further checks whether hourly resolution improves the DART value to offset the greater complexity of a higher-dim.~search. 

For both of the above formulations, the BO is performed directly over the signed block amplitudes
\(
x_j \in [-\bar B,\bar B]^{d},
\) where $d=24$ and $12$ respectively.
Unlike ARBO-DART, there is no longer a predetermined charge or discharge sign to impose on the decision blocks. Consequently, the SoC constraints are piecewise linear in the signed dispatch variables $x_j$ and cannot be supplied directly as linear constraints to \texttt{optimize\_acqf}. 
We therefore optimize the acquisition function using a sampling-based procedure. At each BO iteration, we generate LHS containing \(2^{12}\) candidate profiles over \([-\bar B,\bar B]^d\), remove all profiles that violate the DA power or SoC constraints, and select the candidate that maximizes the UCB criterion \eqref{eq:ucb_final}. We initialize with LHS designs of size \(n^{(12)}_0=20\) and \(n^{(24)}_0=29\), respectively, and then run BO for up to \(n_{\max}=500\) additional RT evaluations, using the same stopping tolerance, \(\varepsilon_{\mathrm{tol}}=0.1\), as ARBO-DART. 

The third comparator is a sequential greedy approach. We first solve the DA-only problem in Section~\ref{sec:da-only} and then pass the resulting DA profile to the RT solver to optimize recourse once. This approach is much computationally faster because the DA and RT decisions are optimized separately. 

\begin{table}[!htb]
\centering
\caption{Economic, computational, and energy decomposition for BO comparators. PnL values are in dollars; aggregate dispatch percentage is from \eqref{eq:cycle} and the RT recourse $L^2$-norm is from \eqref{eq:l2-norm}. Highest PnL in each Case Study is bolded.}
\label{tab:comparator_pnl_decomposition}
\begin{tabular}{rr|rrrr}
\hline
\textbf{CS} &
\shortstack{\textbf{Solver}} &
\shortstack{\textbf{PnL}} &
\shortstack{\textbf{RT solver} \\ \textbf{Evals.}} & 
\shortstack{\textbf{ADP} \\ (\%)} &
\shortstack{\textbf{RT Norm} \\ \(\boldsymbol{\|\Delta^{\mathrm{RT}}\|_2}\)} 
\\ \hline\hline
1 & ARBO-DART  & \textbf{104.50} & 201 & 125 & 1.29 \\
1 & Full BO      & 104.27 & 529$^*$ & 130 & 4.30  \\ 
1 & 12-dim. BO      & 103.95 & 346 & 125 & 4.27  \\ 
1 & Greedy     & 104.15& 1 & 126 & 0.89  \\ 
\hline
4 & ARBO-DART & 140.53 & 102 & 102 & 1.41  \\ 
4 & Full BO      & \textbf{140.67} & 529$^*$ &  118 & 4.56  \\ 
4 & 12-dim. BO      & 140.64 & 464 & 108 & 3.88  \\ 
4 & Greedy     & 139.89 & 1 & 106 & 0.82  \\ 
\hline
2 & ARBO-DART    & \textbf{113.57} &  198 & 133 & 3.46  \\
2 & Greedy     & 103.89 & 1 & 126 & 0.94  \\ 
\hline
\end{tabular}

\end{table}

Table \ref{tab:comparator_pnl_decomposition} shows the head-to-head performance of ARBO-DART against these three alternatives. While the Full 24-dim.~solver achieves comparable PnL (23 cents less than ARBO-DART on CS1, 14 cents more on CS2), it is much slower. In fact, it never reaches the termination criterion, so the reported values are based on what Full found after 500 evaluations, which is multiple times more effort than ARBO-DART. This shows that directly tackling a 24-dim.~black-box optimization is highly inefficient as the search space is too big. The 12-dim.~solver does converge, but still takes about twice as long as ARBO-DART and is 55 cents worse than ARBO-DART on CS1. A key reason is that the adaptive refinement allows us to assign predetermined block signs, reducing the feasible search space by a factor of $2^{d_m}$.

%Overall, ARBO-DART produces DA profiles that are better aligned with the intended DART structure: the DA schedule captures the main arbitrage opportunities, while RT recourse serves as a dynamic corrective mechanism. In contrast, the direct 12- and 24-dimensional BO comparators require substantially larger RT adjustments to offset imperfect DA allocations. The sequential DART method remains a computationally inexpensive and practically relevant comparator, but its performance gap will be large whenever favorable RT price spreads occur outside the charging and discharging windows selected by the DA-only optimization.

\begin{figure}[!htb]
    \centering
    \includegraphics[width=0.6\linewidth]{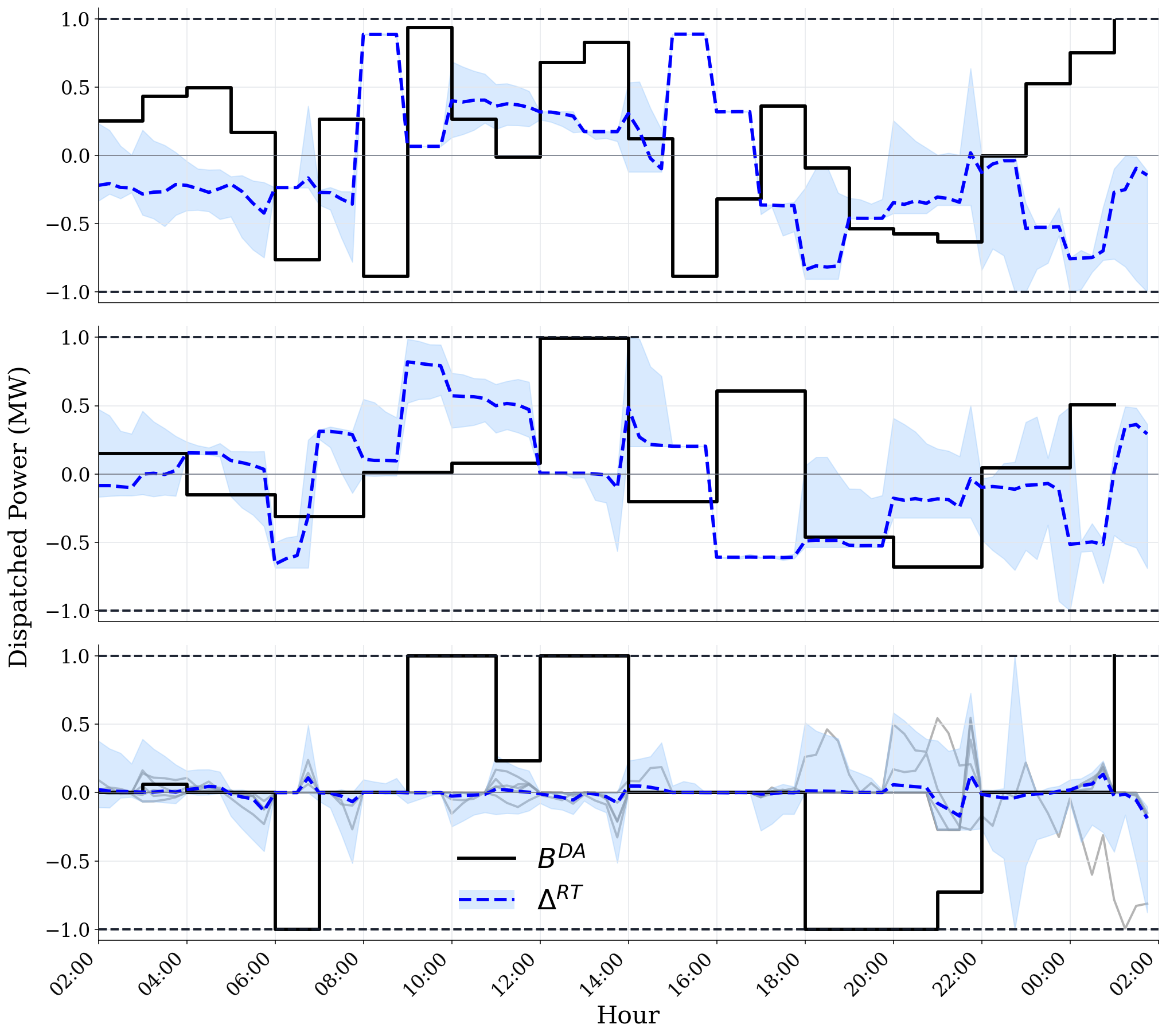}
    \caption{Comparator solutions for Case Study 1. \emph{Top panel:} Full 24-dim.~solution. \emph{Middle:} 12-dim.~fixed-resolution solution; \emph{Bottom:} Greedy DA solution. In each panel, the DA profile is in black, while
    the blue curves show the average RT adjustment along with its $99$-th percentile band. 
   }
    \label{fig:cs1_fixed_resolution_comparators}
\end{figure}

Moreover, as shown in Figure~\ref{fig:cs1_fixed_resolution_comparators}, the solutions provided by the fixed-resolution solvers exhibit multiple hours with offsetting DA and RT actions. Such over- or under-allocations, where RT actions have the reverse sign of the DA position, effectively mean that the optimization did not fully converge. As a result, these comparators do a lot more of RT recourse (see $\| \mathbf{\Delta}^{\mathrm{RT}}\|_2$ in Table \ref{tab:comparator_pnl_decomposition}) which is economically wasteful. 

At the other extreme, the Greedy approach requires only one RT evaluation. It actually outperforms 12-dim.~solver in CS1 but trails behind both 24-dim.~and 12-dim.~solvers in CS4. In both cases, it is worse than ARBO-DART: by $35$ cents and $64$ cents in CS1 and CS4, respectively, highlighting the value of co-optimization. Co-optimization becomes essential when the average DART spread is non-zero. The last row of Table \ref{tab:comparator_pnl_decomposition} shows that effect in Case Study~2 of Section \ref{sec:dart-spread}. Since the DA price profiles match, the Greedy solution of CS2 is the same as that of CS1. In particular, the corresponding DA-only optimal schedule commits to discharge during 20:00--22:00, so that the Greedy solver cannot exploit the higher RT prices during that period and loses nearly \$10 (\$103.89 vs \$113.57) of PnL relative to ARBO-DART. This insensitivity to DART spreads is a major limitation of Greedy.

\subsection{Ablation studies}
\label{sec:ablation}

To assess the sensitivity of ARBO-DART to its algorithmic choices, we conduct a set
of ablation experiments. Returning to Case Study 1, 
Table~\ref{tab:ablation_study} reports the effect of three components: the
random seed used in the BO and RT solvers, the impact of the initial DA partition blocks, the number of cuts $q_m$ and the BO budget at each refinement
stage. Unless stated otherwise, we use the default
initial partition
$\cP^{(1)}=\{[9,13)_{+},[16,20)_{-}\}$ and $q_m=3$. %The reported runtimes are based on parallel runs of single-threaded BO iterations on an AMD Ryzen Threadripper PRO 5965WX 24-Core processor.

\begin{table}[!htb]
\centering
\caption{Ablation study of the ARBO-DART algorithm.}
\label{tab:ablation_study}
\resizebox{\textwidth}{!}{%
\begin{tabular}{cccccccccc}
\hline
\textbf{Seed} &
Initial $\boldsymbol{\cP^{(1)}}$ &
\shortstack{ \textbf{Cuts}\\ $\boldsymbol{q_m}$} &
\shortstack{\textbf{Stages} \\$\boldsymbol{m}$} &
\shortstack{\textbf{Final} \\  $\boldsymbol{d_\mathcal{M}}$} &
\shortstack{\textbf{Final} \\ $\boldsymbol{\mathcal{S}(\mathcal{K}_1^{(\mathcal{M})})}$} &
\shortstack{\textbf{obj. value} \\ $\boldsymbol{y^*(\cP^{(\mathcal{M})})}$} &
\shortstack{\textbf{RT solver} \\ \textbf{Evals.}} &
%\shortstack{\textbf{Runtime} \\ \textbf{(h:min)}} &
\shortstack{\textbf{Non-zero} \\ \textbf{DA blocks}} \\
\hline\hline 
1 & $\{[9,13)_{+},[16,20)_{-}\}$ & 3 & 5 & 12 & 0.042 & 104.46 & 201 %& 3:54
& 9 \\
2 & $\{[9,13)_{+},[16,20)_{-}\}$ & 3 & 5 & 13 & 0.032 & 104.47 & 200 %& 3:53
& 10 \\
3 & $\{[9,13)_{+},[16,20)_{-}\}$ & 3 & 5 & 11 & 0.035 & 104.42 & 194 %& 3:43
& 8 \\
4 & $\{[9,13)_{+},[16,20)_{-}\}$ & 3 & 5 & 12 & 0.043 & 104.45 & 194 %& 3:44 
& 10 \\
\hline
1 & $\{[9,11)_{+},[16,18)_{-}\}$ & 3 & 5 & 13 & 0.239 & 104.37 & 205 %& 3:56
& 10 \\
1 & $\{[9,12)_{+},[16,19)_{-}\}$ & 3 & 5 & 12 & 0.098 & 104.53 & 195 %& 3:45 
& 9 \\
\hline
1 & $\{[9,11)_{+},[16,18)_{-}\}$ & 1 & 9 & 8 & 0.295 & 104.28 & 279 %& 5:04 
& 6 \\
1 & $\{[9,11)_{+},[16,18)_{-}\}$ & 2 & 5 & 9 & 0.364 & 104.42 & 171 %& 3:12
& 7 \\
1 & $\{[9,11)_{+},[16,18)_{-}\}$ & 4 & 5 & 15 & 0.317 & 104.29 & 213 %& 4:10
& 10 \\
\hline
%FB$1$  & $\{[9,11)_{+},[16,18)_{-}\}$ & 3 & 5 & 12 & 0.041 & 104.47 & 210 %& 4:08 
%& 9 \\
%FB$2$  & $\{[9,11)_{+},[16,18)_{-}\}$ & 3 & 5 & 12 & 0.049 & 104.39 & 205 %& 4:08 
%& 9 \\
%\hline
\end{tabular}%
}
\end{table}

The first four rows of Table \ref{tab:ablation_study} examine sensitivity to the BO random seed. The final objective value is stable across seeds, ranging from $104.42$ to $104.47$, a spread of
only $5$ cents. All four runs terminate after $\mathcal{M}=5$ refinement stages, with
BO evaluation counts within $\pm 10$ of each other. The resulting
DA profiles also look very similar modulo some variability in blcok boundaries and magnitudes, cf.~Supplementary Figure~\ref{fig:seed_final_BDA}. 

The next two rows vary the initial partition. Specifically, we change the length of the initial DA charge and discharge blocks. Starting from two-hour and three-hour charge-discharge blocks
produces final DART values of $104.37$ and $104.53$, respectively, compared
with $104.46$ under the four-hour initialization. These results validate that ARBO-DART has limited
sensitivity to the initial $\cP^{(1)}$ length.

The last three rows of Table \ref{tab:ablation_study} vary the number of new blocks $q_m$ added at each refinement
stage. As expected, the most conservative setting $q_m=1$ requires more
refinement stages, terminating after $\mathcal{M}=9$, with substantially longer run-time. For $q_m=2,3,4$, the algorithm
terminates after the same number of refinement stages, with $q_m=2$ needing the
fewest number of RT evaluations. Across all configurations in Table~\ref{tab:ablation_study}, the terminal BO dimension \(d_{\mathcal{M}}\) ranges from \(8\) to \(15\), while the corresponding DA profiles contain only \(6\) to \(10\) nonzero blocks. Thus, even when the refined parameterization is moderately high-dimensional, the resulting DA solution remains sparse.

%Recall that the adaptive scheme allocates \(n_{\max}^{(m)}=n_0^{(m)}+20\) at intermediate stages \(m=1,\ldots,\mathcal{M}-1\) and \(n_{\max}^{(\mathcal{M})}=n_0^{(\mathcal{M})}+5d_{\mathcal{M}}\) at the terminal stage. To assess budget sensitivity, we compare it with a fixed-budget variant that uses \(n_{\max}^{(m)}=n_0^{(m)}+30\) at every stage for seeds 1 and 2. As shown in the final two rows of Table~\ref{tab:ablation_study}, the fixed-budget runs also terminate after \(\mathcal{M}=5\) stages and achieve objective values within approximately one cent for Seed 1 and eight cents for Seed 2 of those obtained by the adaptive scheme. Although the final partitions are similar, their intermediate partitions differ. Figure~\ref{fig:fixed_adaptive_comparison} illustrates this behavior for Seed 1: the fixed-budget variant performs better at stages \(m=3\) and \(m=4\), while the larger terminal-stage budget of the adaptive scheme produces a comparable final objective. 

%\begin{remark}
We also made additional runs of ARBO-DART to check the impact of pruning zero-decision blocks and of the BO budget at intermediate stages. We find that taking $n_{\max}^{(m)}$ too low leads to erroneous cuts which slows down the search, but otherwise the role of $n_{\max}^{(m)}$ is muted. Intuitively, intermediate stages only matter as far as correctly ranking the block scores, and the intermediate PnL values are irrelevant. In contrast, at the terminal stage we care solely about the PnL, hence our choice to set a conservative large budget  $n_{\max}^{(\mathcal{M})}=n_0^{(\mathcal{M})}+5d_{\mathcal{M}}$  to ensure high-accuracy search for the ultimate DA commitment. Figure~\ref{fig:fixed_adaptive_comparison} in the Appendix illustrates that there is scope to speed up the algorithm by fine-tuning these internal parameters. A practical heuristic is to allocate \(20\text{--}30\) BO iterations at each intermediate stage, with additional BO budget for the terminal stage. As far as removing
pruning, we find that without it a quantitatively similar final policy is obtained, however the runtime increases by 10\%+ due to having to run BO over higher
dimensional search spaces in later refinement stages, which makes each \texttt{optimize\_acqf} call more expensive.
%\end{remark}

\section{Conclusion}
\label{sec:conclusion}

This paper developed ARBO-DART, an adaptive BO framework for DART co-optimization of BESS. In our setup, the operator optimizes the DA schedule and then adjusts through closed-loop RT control, accounting for stochastic RT prices and BESS capacity constraints.  We propose to utilize BO for the outer optimization in order to minimize evaluations of the expensive RT stochastic control solver.
Rather than trying to directly optimize the full 24-dim.~DA schedule, our method starts from a low-dimensional
partition and uses the learned RT recourse policy to guide adaptive refinement of the DA decision space. RT recourse acts as a diagnostic for refining DA bidding resolution and also allows to pre-assign charge/discharge directions to each refined block. As we demonstrate, the latter dramatically simplifies the search space, as well as efficiently identifies the effective dimension of the DA schedule, helping ARBO-DART to find a strong solution multiple times faster than full-dimensional BO.

We show that ARBO-DART is effective across different price-curve shapes and DART spread structures and is stable under a range of algorithm initializations. Our solutions also yield economically interpretable DA profiles with a small number of charge and discharge intervals, while non-adaptive BO comparators are less parsimonious and over-rely on RT recourse.  The case studies also demonstrate the value of DART co-optimization over greedy optimization, particularly when the DART spread has a non-zero mean.

Several directions warrant further investigation. From an algorithmic perspective, it could be worthwhile to design a batch BO that reduces wall-clock time through concurrent RT evaluation of candidate DA profiles. From a modeling perspective, future work should investigate alternative RT price simulators and DART bias structures, including the impact of the timing, magnitude (e.g.~heavy tailed rather than Gaussian), and persistence of RT price deviations on optimal DA allocations and subsequent RT recourse. Risk-averse formulations also present a natural direction, where DA positions may be used not only to maximize expected PnL, but also to mitigate variance induced by RT price uncertainty. Even under low RT costs, operators often prefer to partially lock-in their gains via DA commitments. Another extension would to include additional DA products, such as reserve or frequency-regulation capacity. This would enlarge the DA decision vector and modify the power and SoC constraints to account for reserve headroom and
possible RT activation. Finally, as BESS penetration increases, storage operators may transition from being price takers to price makers, their actions impacting either $P^{DA}$ or $P^{RT}$. This motivates game-theoretic formulations of DART bidding that integrate two-level market-equilibrium models, such as \textcite{gu2022market}, with dynamic multi-agent approaches for RT optimization, such as \textcite{ml_rh_hz_paper}.

\noindent \textbf{} 
\printbibliography
\clearpage

\appendix

% ==== Supplementary Numbering ====
\renewcommand{\thesection}{S\arabic{section}}
\renewcommand{\thesubsection}{S\arabic{section}.\arabic{subsection}}
\renewcommand{\thefigure}{S\arabic{figure}}
\renewcommand{\thetable}{S\arabic{table}}
\renewcommand{\theequation}{S\arabic{equation}}
\setcounter{figure}{0}
\setcounter{table}{0}

% --- Title ---
{\bf
Supplementary Material for\\[2pt]
\emph{An Algorithmic Framework for Multi-Timescale Co-Optimization of Battery Storage in Day-Ahead and Real-Time Markets}
}

\section{Additional BO Diagnostics}
\label{app:bo_diagnostics}

\begin{figure}[H]
    \centering
    \includegraphics[width=1\linewidth]{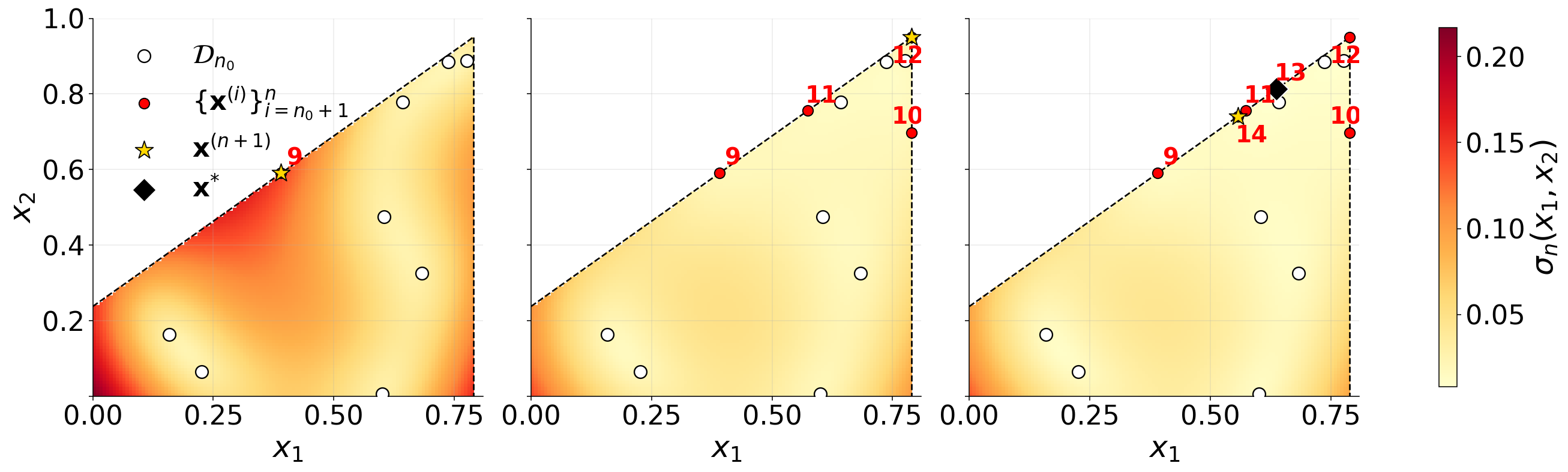}
    \caption{Illustrating DART co-optimization over 2-block parameterization $d=2$. We show the GP posterior standard deviation surfaces \(\sigma_n(x_1,x_2)\) over the feasible polytope \(\mathcal X^{\mathrm{DA}}(\mathcal{P})\) at representative BO iterations $n=8, 11, 13$. White circles denote initial LHS design with $n_0 = 8$, red circles denote $\bx^{(i)}, i=n_0+1,\ldots$, the star denotes the newly proposed candidate $\bx^{(n+1)}$ and the diamond denotes the ultimate $\bx^*$.}
    \label{fig:bo_sigma_panels}
\end{figure}

Supplementary Figure~\ref{fig:bo_sigma_panels} shows the posterior standard deviation surfaces
corresponding to the posterior mean plots in Figure~\ref{fig:bo_mu_panels}.
These panels illustrate the exploration--exploitation behavior of BO through
the spatial pattern of posterior uncertainty $\sigma_n(\bx)$. At \(n=n_0=8\), after only the
initial LHS design points have been evaluated, uncertainty remains
high over regions of the feasible polytope \(\mathcal X^{\mathrm{DA}}(\mathcal{P})\) that are away from the sampled
locations. As additional RT evaluations are added, the posterior uncertainty
contracts, especially around the explored high-value region. In turn, this lowers both the $\mathrm{ucb}$ and $\mathrm{lcb}$ metrics, triggering the BO termination rule.

\begin{figure}[!htb]
    \centering
    \includegraphics[width=0.7\textwidth]{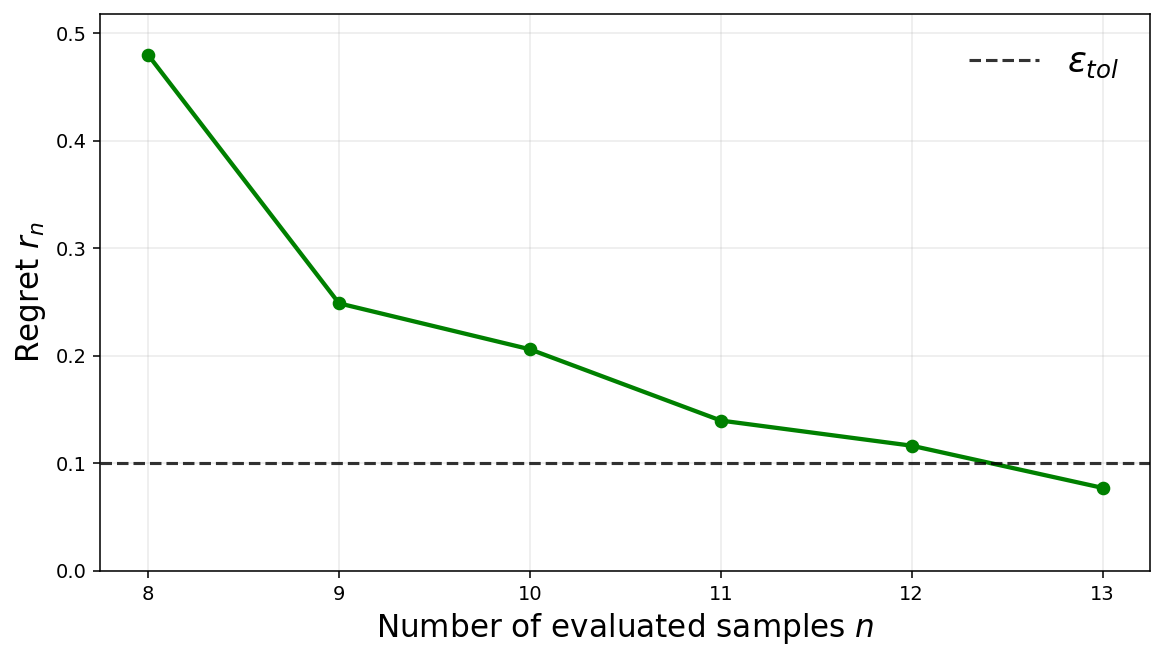}
    \caption{Regret \(\bar r_n\) from \eqref{eq:regret_upper_bound_max} across BO iterations for Case Study 1 at stage $m=1$. The dashed line marks
    the stopping tolerance \(\varepsilon_{\mathrm{tol}}=1\times 10^{-1}\) that triggers the termination rule. }
    \label{fig:bo_regret_bound}
\end{figure}

Supplementary Figure~\ref{fig:bo_regret_bound} plots
the regret upper bound \(\bar r_n\) from the BO stopping rule \eqref{eq:regret_upper_bound_max} across BO
iterations $n$. The regret trends down and falls below the prescribed
tolerance \(\varepsilon_{\mathrm{tol}}=1\times 10^{-1}\) after $n=13$ evaluated
samples, at which point the early termination criterion is triggered.

%%%%%%%%%%%%%%%%%
\section{Representative policies under ARBO-DART}
\label{sec:supp_representative_policies}

\begin{figure}[H]
\centering
\includegraphics[width=0.9\textwidth]{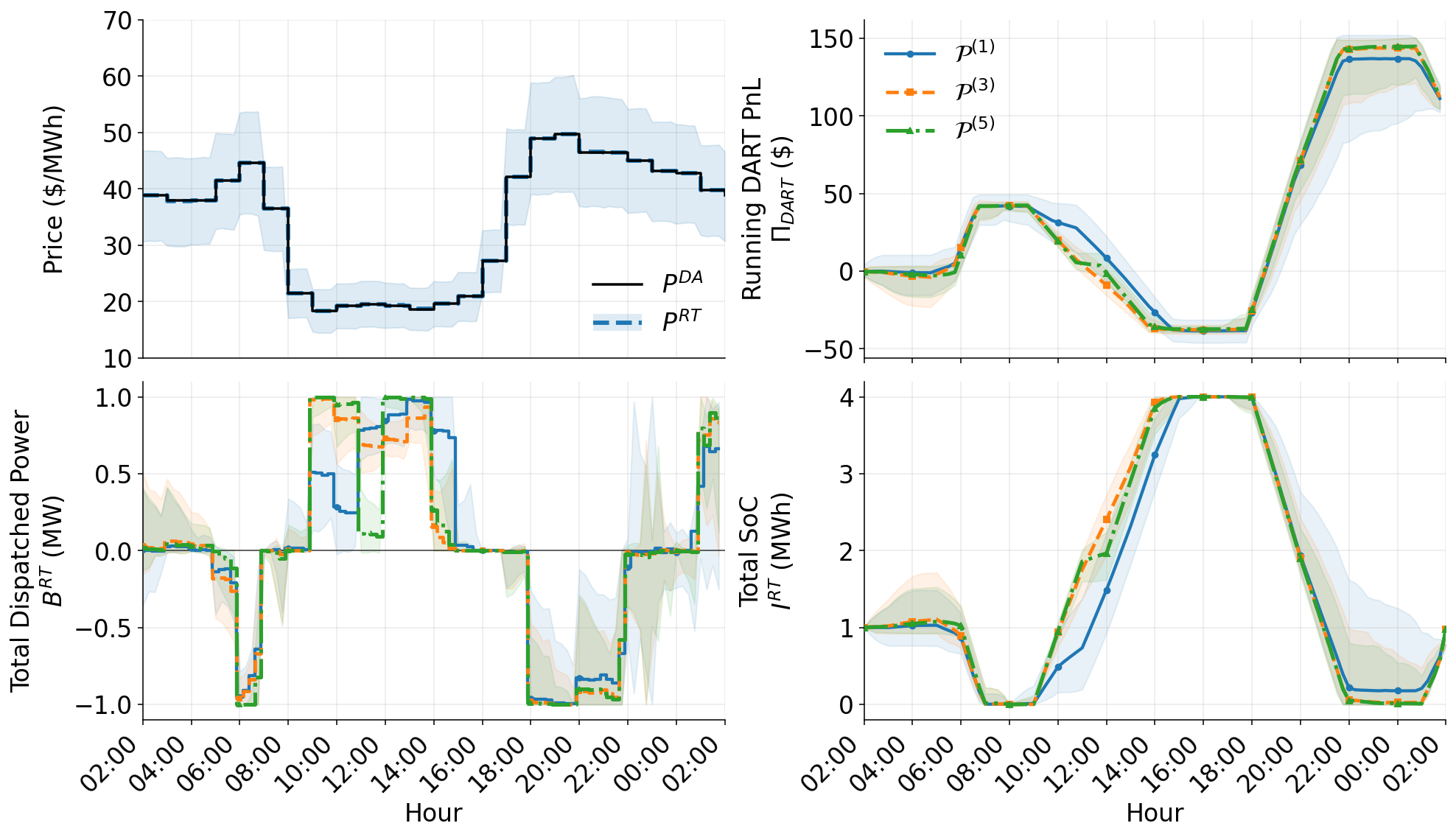}
\caption{Case Study 1: initial partition $\mathcal P^{(1)}$, an intermediate $\mathcal P^{(3)}$, and the final partition $\mathcal P^{(5)}$.
From left to right and top to bottom, the panels show the DA price curve $P^{\mathrm{DA}}$ and expected RT price process, running DART PnL $\Pi^{\mathrm{DA}}$, total dispatched power $B^{\mathrm{RT}}$, and RT SoC $I^{\mathrm{RT}}$. }
\label{fig:supp_three_policies_viz}
\end{figure}

Figure~\ref{fig:supp_three_policies_viz} compares the sequence DART policies generated during representative stages of ARBO-DART algorithm for Case Study 1, corresponding to the initial partition $\mathcal P^{(1)}$, an intermediate partition $\mathcal P^{(3)}$, and the final partition $\mathcal P^{(5)}$. The shaded regions show the 99th-percentile bands across simulated RT paths. As the partition is refined, the policy gains additional flexibility to allocate charging and discharging across the day. This is visible in both the running PnL and SoC trajectories: later refinements track the main arbitrage opportunities more closely and produce sharper PnL accumulation around the evening price peak. 
The optimized objective value increases from $102.95$ under $\mathcal P^{(1)}$ to $104.00$ under $\mathcal P^{(3)}$ and $104.46$ under $\mathcal P^{(5)}$, showing the improvement obtained through adaptive refinement. The refinement also narrows the uncertainty bands; see the running DART PnL and total SoC panels. This reflects a shift from relying on RT recourse toward putting more charge--discharge actions in the DA layer. 

\section{Sensitivity to RT Running Penalty}
\label{sec:supp_gamma_sensitivity}

\begin{figure}[!htb]
    \centering
    \includegraphics[width=0.8\linewidth]{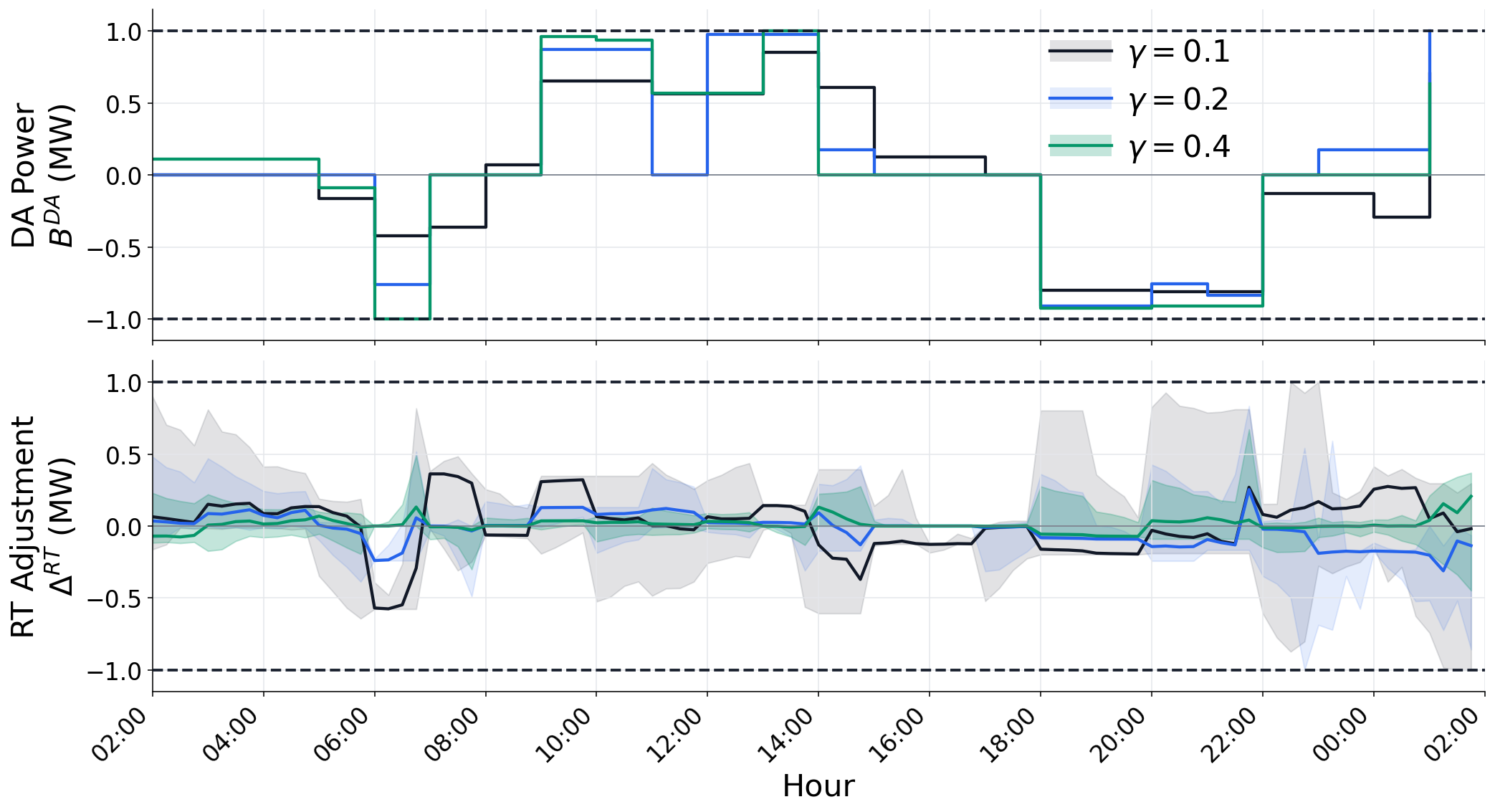}
    \caption{Case Study 1 under different RT
    cost parameters \(\gamma \in \{0.1,0.2,0.4\}\). \emph{Top panel:} Final DA profile output from ARBO-DART. \emph{Bottom panel:} RT recourse
    process and respective 99\% quantile bands. The dashed horizontal lines indicate maximum charge/discharge limits $\pm\bar{B}$. }
    \label{fig:supp_gamma_variation}
\end{figure}

We also examine the sensitivity of ARBO-DART solution in Case Study~1 to the RT
running-penalty parameter \(\gamma\). This comparison
isolates how the penalty on RT recourse changes the balance between DA
commitment and RT flexibility. Figure~\ref{fig:supp_gamma_variation} shows the optimized DA profiles and
expected RT recourse adjustments from ARBO-DART for
\(\gamma \in \{0.1,0.2,0.4\}\). As \(\gamma\) increases, RT adjustments become
more costly, so the solution shifts more of the charging into
the DA schedules. Equivalently, larger values of \(\gamma\) reduce the flexibility
available to the RT policy and lead to smaller expected RT recourse adjustments.
We observe that the main structure is preserved: for all values of $\gamma$, the co-optimized policy charges during the lower-price daytime period and discharges
during the early morning and evening high-price windows. 

%%%%%%%%%%%%%%%%%

\section{Ablation Studies Supplementary}
\begin{figure}[!htb]
    \centering
    \includegraphics[width=\linewidth]{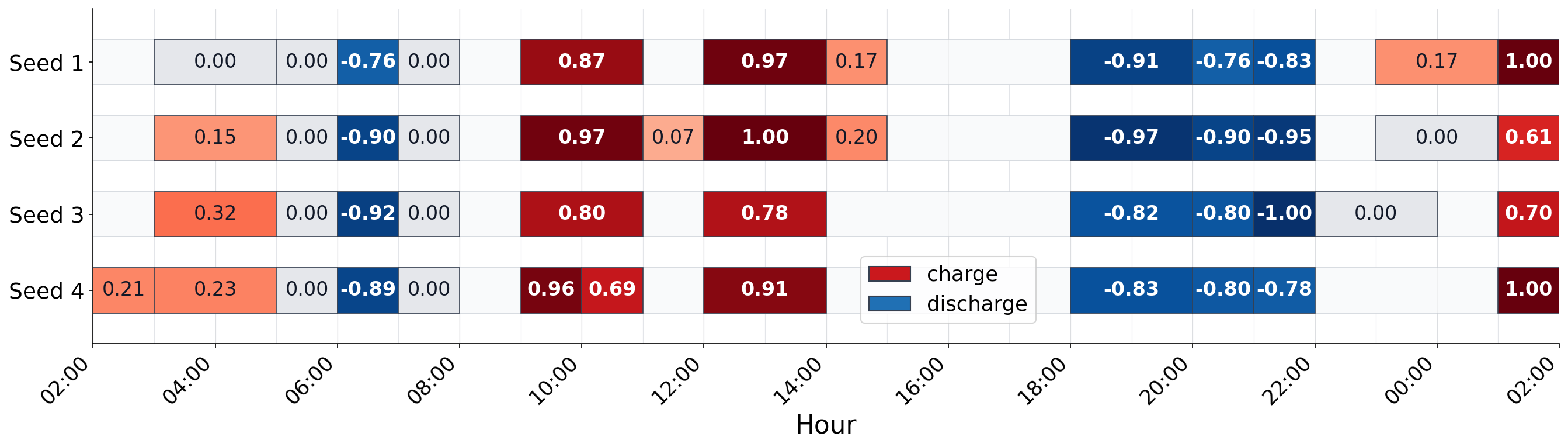}
    \caption{Final DA profiles obtained by ARBO-DART algorithm
    under four different random seeds using the initial partition
    \(\mathcal{P}^{(1)}=\{[9,13)_{+},[16,20)_{-}\}\) and \(q_m=3\) for Case Study 1. Red/blue blocks
    denote charging/discharging $B^{\mathrm{DA}}_h>0$/$B^{\mathrm{DA}}_h<0$.}
    \label{fig:seed_final_BDA}
\end{figure}

Supplementary Figure~\ref{fig:seed_final_BDA} compares the final DA profiles for Case Study 1 obtained under different ARBO-DART random seeds using the default configuration with initial partition $\mathcal{P}^{(1)}=\{[9,13)_{+},[16,20)_{-}\}$  and \(q_m=3\). Across all seeds, the final profiles share the same broad structure: charging is concentrated during the low-price daytime window from 9:00--15:00, followed by DA discharging during the evening peak from 18:00--22:00. We observe that some solvers suggest to charge in the early morning, as well as possibly during 14:00-15:00. There is also some variability regarding the DA charging magnitude during the last 1:00-2:00 hour.

\begin{figure}[!htb]
    \centering
    \begin{subfigure}[t]{0.58\textwidth}
        \centering
        \includegraphics[width=\linewidth]{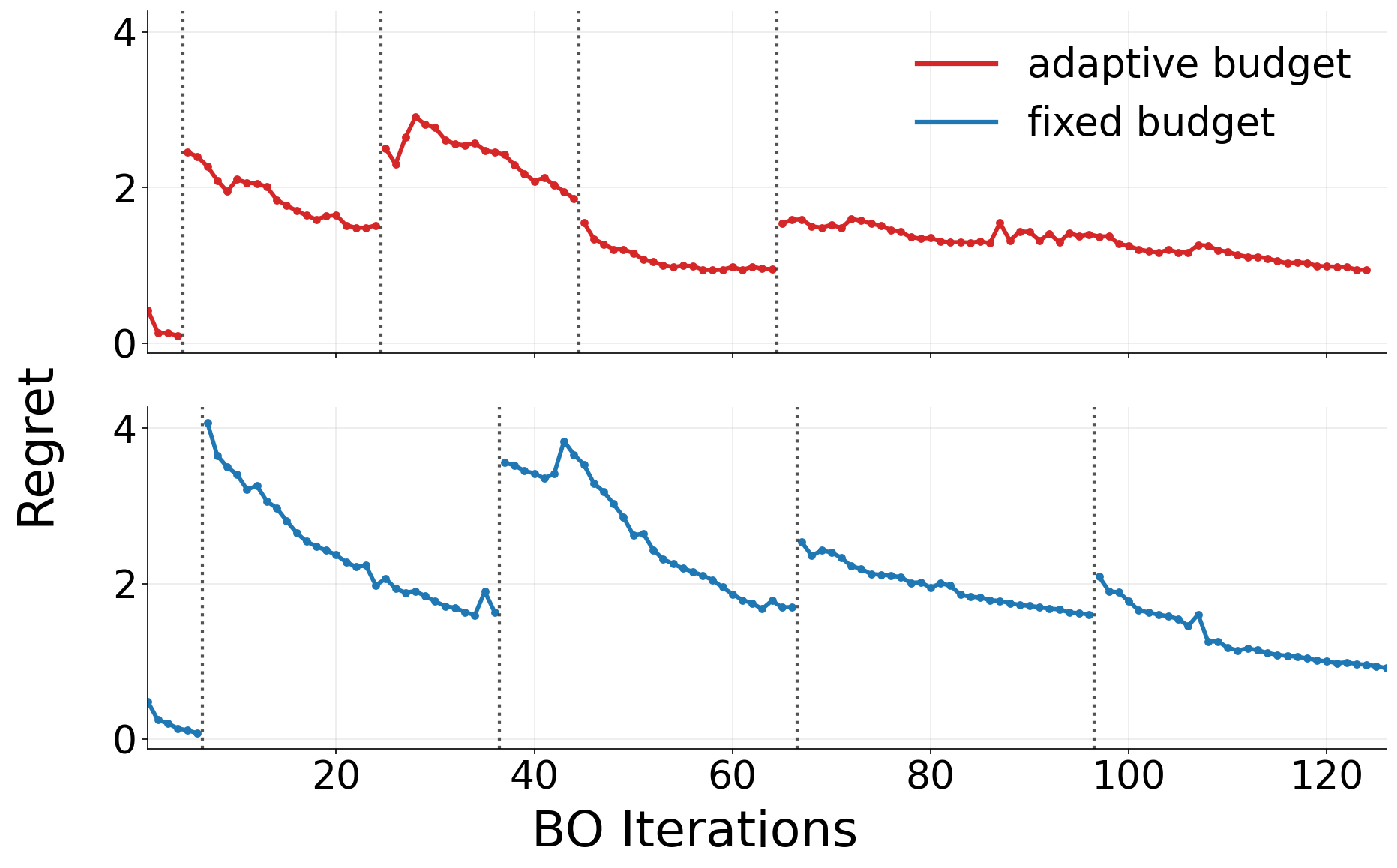}
        \caption{Trace of regret $\bar{r}_n$ within each refinement stage.}
        \label{fig:fixed_adaptive_bo}
    \end{subfigure}
    \hfill
    \begin{subfigure}[t]{0.38\textwidth}
        \centering
        \includegraphics[width=\linewidth]{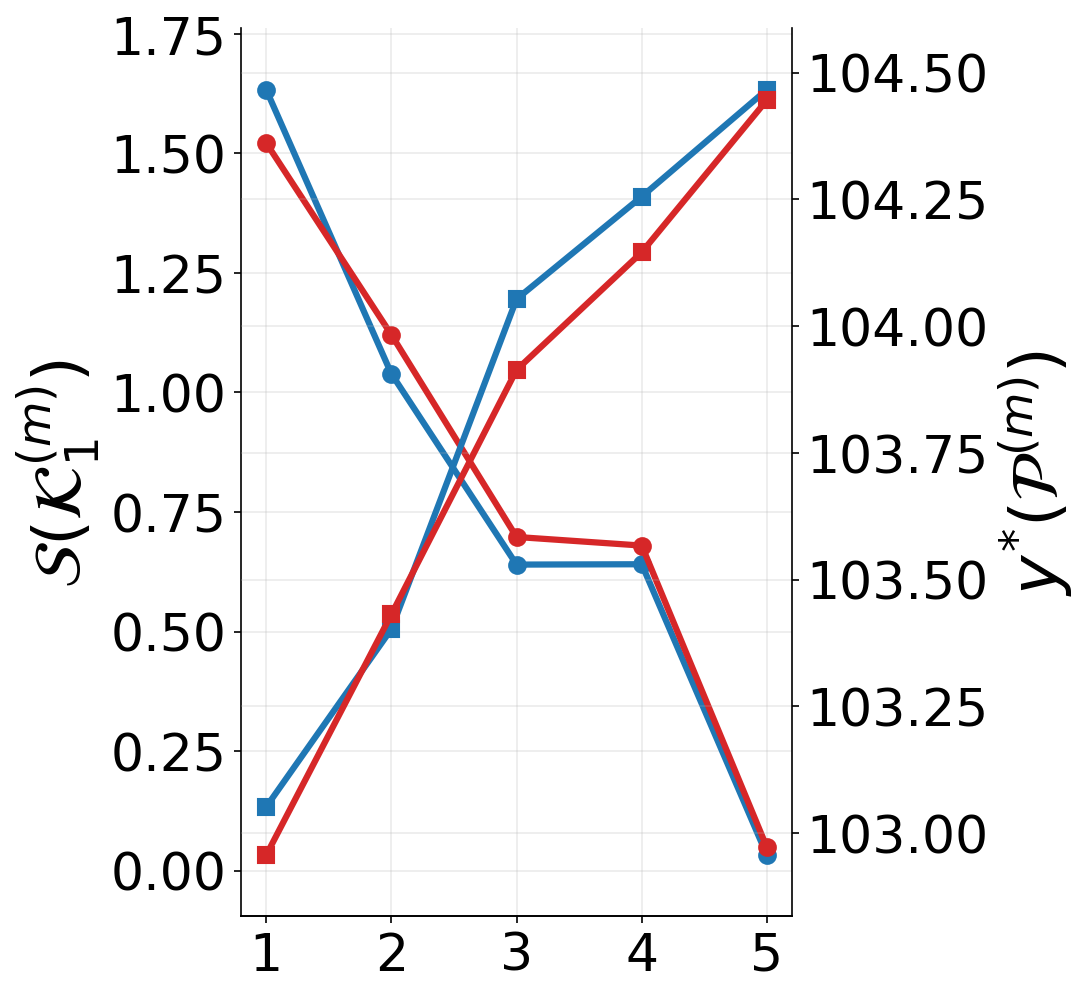}
        \caption{Score and objective value.}
        \label{fig:fixed_adaptive_score}
    \end{subfigure}
\caption{Comparison of schemes with adaptive and fixed BO budgets across refinement
stages. The left panel shows the BO regret traces, with vertical dotted lines
separating refinement stages. The right panel compares the refinement scores
\(\mathcal{S}(\mathcal{K}_1^{(m)})\) and optimized objective values
\(y^{*}(\mathcal{P}^{(m)})\). Both variants terminate after $m=5$
stages and attain similar terminal DART objective values. Results correspond
to Case Study~1 with initial partition
\(\mathcal{P}^{(1)}=\{[9,13)_{+},[16,20)_{-}\}\) and \(q_m=3\).}
    \label{fig:fixed_adaptive_comparison}
\end{figure}

Supplementary Figure~\ref{fig:fixed_adaptive_comparison} compares the default BO budget allocation with a fixed-budget variant for Seed 1. Recall that the  adaptive scheme allocates \(n_{\max}^{(m)}=n_0^{(m)}+20\) at intermediate stages \(m=1,\ldots,\mathcal{M}-1\) and \(n_{\max}^{(\mathcal{M})}=n_0^{(\mathcal{M})}+5d_{\mathcal{M}}\) at the terminal stage. To assess budget sensitivity, we compare it with a fixed-budget variant that uses \(n_{\max}^{(m)}=n_0^{(m)}+30\) at every stage. Both variants terminate after $\mathcal{M}=5$  refinement stages and take $124$ and $126$ BO iterations, see left panel. Though the final values differ just by 1 cent and the final partitions are similar, their intermediate partitions differ. The fixed-budget variant performs better at stages \(m=3\) and \(m=4\), while the larger terminal-stage budget of the adaptive scheme (which uses over 60 evaluations for $m=5$ stage) produces a comparable final objective, see the right panel.

\end{document}